\documentclass[preprint,12pt,3p,authoryear]{elsarticle}

\usepackage{amsmath,amssymb,amsfonts}
\usepackage{bm}
\usepackage{graphicx}
\usepackage{booktabs}
\usepackage{algorithm}
\usepackage{algpseudocode}
\usepackage{lineno}
\usepackage{xcolor}
\usepackage{url}
\usepackage[normalem]{ulem}

\newcommand{\rev}[1]{{#1}}

\graphicspath{{figures/}}
\journal{Journal of the Mechanics and Physics of Solids}

\DeclareMathOperator*{\argmin}{arg\,min}

\begin{document}

\begin{frontmatter}

\title{Autonomous solids with self-determined topology, nonreciprocity and mechanical activity}

\author[stmu]{Mohamed Shaat\corref{cor1}}
\ead{mibrahim1@stmarytx.edu}
\cortext[cor1]{Corresponding author.}

\address[stmu]{Department of Mechanical and Industrial Engineering, St.\ Mary's University,
San Antonio, TX 78228-0337, USA}

\begin{abstract}
Living collectives, such as bird flocks, coordinate their behavior through decisions made locally by their individual members, giving rise to coherent global organization and autonomous adaptation without centralized control. Engineering materials, by contrast, remain passive recipients of decisions made elsewhere: their constitutive behavior is fixed at synthesis, prescribed by external stimuli, or dictated by a controller. Endowing a solid with genuine autonomy has remained elusive because decentralized decision-making processes are inherently stochastic, whereas materials require deterministic, reproducible constitutive laws. Here, we develop an autonomous solid in which identical active particles embedded in an elastic lattice sense their local mechanical state, exchange information with nearby particles, and independently decide whether to relocate to a vacant lattice site. We demonstrate that through decentralized coordination, the material particles collectively reach a consensus on the same internal topology -- that is, the same arrangement of occupied sites -- under the same loading conditions, so the mechanical response remains deterministic despite stochasticity at the level of individual particles. As the autonomous material deforms, it spontaneously reorganizes its topology and its elastic bond network, autonomously modulating its stiffness and switching between achiral and chiral elastic responses. Because these rearrangements depend on the direction of loading, the material responds differently to opposite deformations, exhibiting mechanical nonreciprocity of new origin. Moreover, the dependence of topology evolution on loading history offers a new route to adaptive mechanical activity, where the autonomous material can exchange net mechanical work with its surroundings over closed deformation cycles, acting as either an energy source or an energy sink. These findings establish autonomy as a design principle for solids whose material laws emerge from the local decisions of their constituent units.

\end{abstract}

\begin{keyword}
Autonomous materials \sep Decentralized decision-making \sep Distributed consensus \sep Mechanical metamaterials \sep Stiffness modulation \sep Nonreciprocal elasticity \sep Active solids
\end{keyword}

\end{frontmatter}


\section{Introduction}
\label{sec:1}

Collective behaviors in biological systems are very often governed by decentralized coordination among individual units of the system~\rev{\citep{toner1998,parrish2002,vicsek1995,lukeman2010,vicsek2012}}. An example of this behavior is \textit{murmuration}, where thousands of starlings move in a highly coordinated evolving formation that expands, contracts and twists in apparent unison, with no bird directing the collective motion~\rev{\citep{vicsek2012,shimoyama1996,young2013}}. This flocking behavior emerges from a set of local rules followed by each bird -- align velocity with neighbors, avoid collision, maintain cohesion -- from which a coordinated and adaptive global response emerges in real time without centralized control~\rev{\citep{toner1998,parrish2002,lukeman2010,reynolds1987,olfatisaber2006,partridge1982,deneubourg1989,ruizmirazo2004,anderson2023,anderson2022}}. From an engineering standpoint, such a flock is an \emph{autonomous active substance}, one that produces coherent adaptive global responses through \textit{decentralized} processing of information that is communicated only locally among its constituents~\rev{\citep{olfatisaber2006,olfatisaber2003,ruizmirazo2004,merindol2017}}.

This process bears a close analogy to elastic wave and stress propagation in solids. When a solid is subjected to a boundary load, stresses and elastic waves propagate inward from the loaded surface, \rev{transmitting} mechanical information through tractions and displacements across the material's building blocks, a process governed by the balance of linear and angular momenta~\rev{\citep{mindlin1962,toupin1962,shaat2025a}}. In a murmuration, an analogous information front propagates inward: birds on the boundary detect a threat, adjust their velocities and directions, and this kinematic information is relayed neighbor-to-neighbor, prompting the entire flock to adjust its formation. 

Although both systems coordinate global structural reconfiguration through local, boundary-to-interior information propagation, they differ fundamentally in what happens once the information arrives at an interior unit. In a conventional solid, each unit cell responds according to a fixed, prescribed constitutive law, with its behavior determined entirely by microstructure and material properties established at synthesis~\rev{\citep{hill1963,olson1997,gibson1997,shaat2020e,shaat2018c,shaat2025a}}. Active and stimulus-responsive materials possess adaptive capabilities, yet their unit cells remain passive recipients of information, responding to externally applied stimuli through fixed stimulus--response relationships~\rev{\citep{sun2012,otsuka1999,brandenbourger2019,yang2025,chen2026}}. Programmable and reconfigurable metamaterials can switch between different configurations, but the selection among these configurations is externally controlled~\rev{\citep{cui2014,florijn2014,haghpanah2016,bao2020,fang2022,gregg2024,zhou2026a}}. Across these paradigms, the material is subservient to decisions made externally to its constituent units -- whether those decisions are encoded at synthesis, delivered through external stimuli, or issued by a controller or a feedback law designed to realize a prescribed response. In biological collectives such as murmurations, by contrast, each unit is an \textit{autonomous agent} that independently \emph{decides} how to respond by evaluating the information it receives against its own local rules. It is precisely this capacity for \textit{decentralized autonomous decision-making} at the level of individual units -- absent from existing solid material paradigms -- that the present work translates into synthetic solids. The effort belongs to a broader surge of research that brings the organizational principles of living systems into the scope of materials science and condensed matter physics~\rev{\citep{zhang2021a,correll2014,shankar2022,coulais2021,rubenstein2014a,palacci2013,fruchart2021,zhang2022a,liu2025}} and exploits pathways toward autonomous material systems that can self-organize and self-adapt only through decentralized interactions among their building blocks~\rev{\citep{aubin2024,pfeifer2007,hagan2016,tan2021,markvicka2018,orrego2020,elhelou2024,baulin2026}}. 

A key challenge confronting translating autonomy in the sense described above into synthetic materials is that autonomy in biological systems is often stochastic -- meaning that, under nominally identical conditions, the decentralized interactions between the individual units may produce multiple admissible global responses~\rev{\citep{okubo1986,reynolds2022,reynolds2023,dong2020,olaighleis2018,nguyen2026,moreno2015,kumar2008,adar2004,julius2008,losick2008,zechner2020}}. When directly translated into a solid, such behavior would undermine constitutive predictability and prevent the establishment of a material law altogether. This is exactly why robotic swarms, which achieve autonomy through decentralized coordination, exhibit stochastic collective behavior that is indescribable by a constitutive relationship~\rev{\citep{fruchart2021,zhang2024,goldman2024,meng2024,wang2024,miskin2026}}. This raises the question that motivates the present study: \emph{can a solid be endowed with genuine decentralized autonomy while retaining deterministic constitutive behavior, and if so, under what conditions?} 

Here, we develop a new class of autonomous synthetic solids whose global constitutive behavior is deterministic and yet emergent from a set of decentralized rules followed independently by each building unit of the solid. The autonomous solid considered here is a 2D phononic metamaterial whose unit cell is an elastic lattice partially occupied by identical material particles, each of which is an autonomous mechanical agent capable of: (i) \textit{local sensing} – awareness of its own mechanical state and those of its immediate neighbors; (ii) \textit{independent decision-making} – ability to optimize and \rev{adapt} based on a set of interaction rules; (iii) \textit{local actuation} – ability to migrate and establish connectivity at an optimal lattice site -- which all together define the solid's local decentralized rules. The global constitutive response of the autonomous solid metamaterial arises as a collective emergent property of these local autonomous processes, which are coordinated among particles through a purely decentralized process and without any central control. 

We first introduce the autonomous metamaterial model and elucidate how each particle independently determines its optimal lattice site by executing an assignment problem that minimizes a cost function associated with occupying a vacant lattice site or exchanging sites with another particle. This cost function is assembled exclusively from local mechanical information, namely, the Euclidean distance between the particle's deformed position and candidate lattice sites. We then describe the decentralized coordination through which the particles reach a global consensus on the optimal assignment, driving the metamaterial to autonomously evolve its internal topology, where each completed reassignment constitutes a spontaneous topology-evolution event that changes the material's effective properties. We further identify the mechanism by which the emergent topology, and hence the constitutive behavior, remain deterministic despite stochasticity at the individual particle level. Finally, we establish how topology evolution drives autonomous stiffness modulation, enables adaptive nonreciprocal elastic behavior, and provides a new mechanism for mechanical activity through which the metamaterial can act as either a source or a sink of mechanical work. 

\section{Autonomous Material Model}
\label{sec:2}
\begin{figure}[t]
\centering
\includegraphics[width=\textwidth]{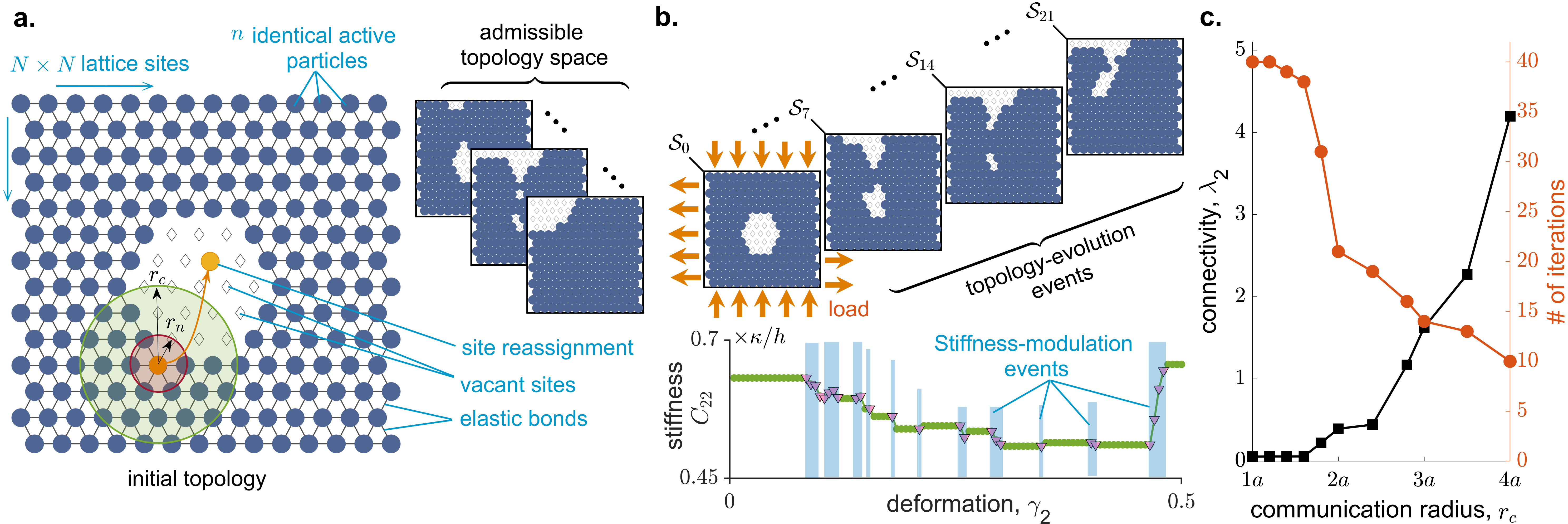}
\caption{\rev{\textbf{Autonomous solid metamaterial.} \textbf{a.} Unit cell of the autonomous metamaterial in its initial topology $\mathcal{S}_0$. Each particle forms elastic bonds with particles within a neighbor radius $r_n$, communicates with particles within a communication radius $r_c$, and can reassign itself to a vacant lattice site. The admissible topology space comprises all possible occupancies of the $N\times N$ lattice sites by the $n$ particles, with representative topologies shown on the right. \textbf{b.} Sequence of topology-evolution events, $\mathcal{S}_0\to\mathcal{S}_1\to\cdots\to\mathcal{S}_m$, triggered as the metamaterial undergoes shear~1 deformation, $\gamma_2$ (see Eq.~\eqref{eq:6}), and the corresponding modulation of the stiffness modulus $C_{22}$. Markers indicate topology-change events, each associated with a change in stiffness. \textbf{c.} Algebraic connectivity $\lambda_2$ of the communication network and number of coordination iterations $t$ required for Algorithm~II to converge, as functions of the communication radius $r_c$. Stiffness is normalized by $\kappa/h$ and lengths by the lattice spacing $a$, where $\kappa$ is the elastic bond stiffness and $h$ is the out-of-plane thickness (see~\ref{app:A}).}}
\label{fig:F1}
\end{figure}

The autonomous solid material considered here is a phononic metamaterial whose unit cell is an elastic domain containing $N\times N$ lattice sites, partially occupied by $n<N^{2}$ active, identical material particles (Fig.~\ref{fig:F1}a). The \textit{admissible topology space} $\{\mathcal{S}_0,\mathcal{S}_1,...\}^T$ of the autonomous metamaterial is determined by the combinatorics of how the $n$ particles can occupy the $N^2$ lattice sites without having two particles sharing the same site. Each particle occupies an initial lattice site -- assigned through a prescribed initial topology $\mathcal{S}_0$ -- and establishes elastic bonds with the neighboring particles that fall within a predefined neighbor zone of radius $r_{n}$\rev{; the host lattice, the initial topology and the elastic bond network used in the present computations are detailed in~\ref{app:A1}}. Unlike conventional solids, in which the internal topology is fixed at synthesis, the metamaterial permits its particles to detach from their current sites and relocate to new ones as the domain deforms, driving the metamaterial through a \emph{topology-evolution event} $\mathcal{S}_m$ (with $m>0$) by which its effective properties change (Fig.~\ref{fig:F1}b).

Each particle is an autonomous mechanical agent endowed with four capabilities, which collectively constitute the \textit{local rules} of the autonomous metamaterial: (i) \emph{sensing} -- each particle is aware of its current position, its displacement, and of the neighboring particles and vacant sites within its neighbor zone $r_{n}$; (ii) \emph{elastic bonding} -- the particle establishes elastic bonds with all particles within the neighbor zone $r_{n}$; (iii) \emph{wireless communication} -- each particle computes a set of \textit{cost values} for occupying the lattice sites located within a communication zone of radius $r_{c}$ (in general $r_{c}\geq r_{n}$), and broadcasts this cost information to the particles within that zone; (iv) \emph{autonomous site reassignment} -- the particle executes a
reassignment process based on the cost information communicated, adopting a decentralized assignment algorithm, and upon this it relocates to a new site and establishes new elastic bonds and communication links, thereby changing the material topology. It is this capacity for spontaneous topology evolution -- driven here by the autonomous decisions of individual particles \rev{rather than} by an external controller -- that separates this autonomous material from all existing material paradigms.

\subsection{Assignment Problem}
The reassignment of particles among the admissible lattice sites is formulated as a constrained assignment problem that is independently executed by each particle:
\begin{equation}
\min_{\pi}\;\sum_{i=1}^{n}c_{i\pi(i)}
\qquad\text{subject to}\qquad \pi : \{1,...,n\}\to \mathcal{L},\qquad\pi(i)\neq\pi(k)\quad\forall\, i\neq k ,
\label{eq:1}
\end{equation}
\noindent where $\pi$ is an assignment vector for the $n$ particles, $\mathcal{L}$ denotes the set of $N^2$ lattice sites, $\pi(i)$ is the lattice site assignment for particle $i$, and $c_{i\pi(i)}$ is the cost associated with assigning particle $i$ to lattice site $\pi(i)$. In the present work, the cost value is defined as the Euclidean distance from the particle's current position $\textbf{x}_i$ to the position of a candidate site $\textbf{X}_j$:
\begin{equation}
c_{ij}=\|\textbf{x}_i-\textbf{X}_j\|.
\label{eq:2}
\end{equation}

The assignment process is triggered by deformation. As the elastic domain deforms, particles are displaced from their reference positions, thereby changing their distances to the admissible lattice sites. Each particle therefore re-evaluates the cost values for occupying candidate sites and, together with the communicated cost information, independently formulates and executes the assignment problem in Eq.~(\ref{eq:1}) to identify its optimal lattice site. The independent execution of this autonomous optimization process by each particle can ultimately drive the metamaterial to change its topology \rev{ (see~\ref{app:A2} for the implementation of this deformation-triggered reassignment)}. 

Two key features set the proposed topology-evolution process \rev{apart} from conventional \textit{topology optimization}, in which an optimal material topology is sought to achieve a prescribed objective function~\rev{\citep{sigmund2001,sigmund2013}}. First, the cost function in the proposed autonomous optimization process is defined exclusively in terms of local information, e.g., Euclidean distance, accessible to each particle from its own sensing and communication capabilities alone. This locality is a defining feature of the autonomous metamaterial, and it is precisely what makes the optimization process \textit{decentralized} and \textit{rule-driven} rather than externally programmed or controlled. This stands in contrast to conventional topology optimization, which constructs its objective function from macroscopic, global material properties that require knowledge of the full system response. In the autonomous metamaterial, no central optimizer constructs or solves the assignment problem; each particle independently formulates and executes its own assignment problem based on information available through local sensing and communication. Second, the cost is evaluated on the \emph{current} deformed state, so that the optimization target itself evolves with the material state; consequently, topology evolution and deformation are intrinsically coupled.
 
Although the cost function encodes only local fields and each particle independently executes the assignment problem, the process \rev{can still} produce a globally unique optimum topology through decentralized coordination among the material particles: each particle broadcasts its local cost values to its immediate neighbors, neighbors propagate this information through the communication network hop by hop, and these purely local exchanges are aggregated into a global consensus among all particles. Thus, the global optimum is not prescribed from above, as in conventional topology optimization methods, but assembled from below through a decentralized coordination process. This bottom-up consensus mechanism is what connects the locality of the cost function to the global uniqueness of the optimal topology.

\subsection{Decentralized Coordination}
The assignment problem in Eq.~(\ref{eq:1}) is solved here adopting the Hungarian algorithm as the optimization method (see Algorithm~I)~\rev{\citep{ismail2017,kuhn1955,jonker1986,millstettey2007,sui2017,chopra2017,samiei2019,kume2021,huang2025}}. This algorithm is executed independently by each material particle, and, when supplied with the complete cost matrix $\textbf{c}$, it returns an optimal assignment of all particles to lattice sites that minimizes the total assignment cost. However, no particle initially possesses the complete cost matrix. Each particle senses and communicates only within a finite zone and therefore initially has access to a partial set of cost values. For the material particles to reach a global consensus, formulate the same assignment problem, and thereby arrive at the same optimal assignment (topology), a decentralized coordination procedure is adopted, as illustrated in Algorithm~II.

Algorithm~I operates on a received cost matrix $\textbf{c}$ by successive row and column reductions, which leave the identity of the optimal assignment unchanged while exposing a set of zero entries. The test for the existence of an \rev{optimal} assignment $\pi$ is that the minimum number of lines needed to cover all zero entries in the cost matrix equals the number of particles. When fewer lines are sufficient, the cost matrix is further reduced and the test is repeated until the optimal assignment can be identified. 
\begin{algorithm}[t]
\caption{Hungarian algorithm}
\label{alg:hungarian}
\begin{algorithmic}[1]
\Require Cost matrix $\textbf{c}$ with scalar entries $c_{ij}$.
\State \textbf{Row reduction:} $c_{ij}\leftarrow c_{ij}-s_{i}$, where $s_{i}$
       is the smallest entry of row $i$.
\State \textbf{Column reduction:} $c_{ij}\leftarrow c_{ij}-s_{j}$, where
       $s_{j}$ is the smallest entry of column $j$.
\State Find the minimum number $\bar{n}$ of lines, drawn through rows and
       columns, that cover all zero entries of $\textbf{c}$.
\If{$\bar{n}=n$}
  \State An optimal assignment exists: 
         $\pi(i)=\argmin_{j}\left(c_{ij}\right)$. \Return $\pi$.
\Else
  \State Identify the smallest uncovered entry of $\textbf{c}$ $\hat{s}$.
  \State Subtract $\hat{s}$ from every uncovered row entry. 
  \State Add $\hat{s}$ to every covered column entry.
  \State \textbf{go to} 3.
\EndIf
\end{algorithmic}
\end{algorithm}

The optimality of the assignment $\pi$ obtained from Algorithm~I and the global consensus among all particles depend on the cost matrix supplied to the algorithm. To make this precise, let $\mathcal{J}_{i}$ denote the set of lattice sites within the communication radius $r_{c}$ of particle $i$, and let
$\mathcal{N}_{i}$ denote the set of particles within that communication zone. Each particle $i$ maintains a local cost matrix $\textbf{c}^{i}$ of dimension
$n\times N^{2}$, initialized as
\begin{equation}
c^{i}_{rj}=
\begin{cases}
\|\textbf{x}_r-\textbf{X}_j\|, & r=i \ \ \text{and} \ \ j\in\mathcal{J}_{i},\\[3pt]
\infty, & \text{otherwise},
\end{cases}
\label{eq:3}
\end{equation}
so that particle $i$ initially knows only its own cost values for occupying lattice sites within its communication zone. Applying Algorithm~I to Eq. (\ref{eq:3}) yields a provisional assignment $\pi^i$, but this assignment does not account for the claims of particles outside the communication zone. Consequently, two particles outside each other's communication zones may provisionally claim the same site, thereby violating the one-to-one assignment constraint in Eq.~(\ref{eq:1}). 

Such conflicts are resolved, and global consensus is achieved, through the decentralized propagation of cost information described in Algorithm~II. Each particle $i$ receives the cost information from its communication neighbors $k\in \mathcal{N}_i$, based on which it updates its local cost matrix $\textbf{c}^i$. Through iterative neighbor communication and cost information updates, each particle progressively assembles a richer cost matrix, and as this information propagates throughout the communication network, the local cost matrices of all particles converge to the same global, complete cost matrix. At this point, each particle independently executes Algorithm~I using the \textit{same} global cost matrix, solving the same assignment problem in Eq.~(\ref{eq:1}), and obtains the same optimal assignment \(\pi\). While the optimization process itself is independently executed by each material particle, the resulting optimal assignment is globally consistent and conflict-free. \rev{The synchronous implementation of Algorithm~II, its convergence criterion, and the exact assignment solver used in the present computations are detailed in~\ref{app:A2}.} 
\begin{algorithm}[t]
\caption{Decentralized coordination}
\label{alg:decentral}
\begin{algorithmic}[1]
\Require Communication radius $r_{c}$ and current particle positions.
\For{each particle $i$ in parallel}
  \State Form the local cost matrix $\textbf{c}^{i}$ with entries $c_{rj}^i$ defined in Eq.~(\ref{eq:3}). 
  \State Apply Algorithm~I to $\textbf{c}^{i}$ to obtain a provisional assignment $\pi^{i}$.
\EndFor
\For {iteration $t=1,2,...$ until $\pi^1=\pi^2=...=\pi^n=\pi$ during an iteration}
  \For{each particle $i$ in parallel}
    \For {each neighbor $k\in\mathcal{N}_{i}$} 
    	\State Connect with neighbor $k$ and receive $\textbf{c}^{k}$.
    	\State Update $\textbf{c}^i$: \textbf{if} $c^{k}_{rj}\neq\infty$ \textbf{and} $r\ne i$ \textbf{then} $c^{i}_{rj}\leftarrow c^{k}_{rj}$.
        \State Update row $i$ in $\textbf{c}^i$ based on information received from neighbor $k$.
    	\State Apply Algorithm~I to $\textbf{c}^{i}$ to obtain a new provisional
           assignment $\pi^{i}$.
    \EndFor
  \EndFor
\EndFor 
\State \Return the conflict-free assignment $\pi$.
\end{algorithmic}
\end{algorithm}

The performance of the decentralized coordination process in Algorithm~II -- whether and how rapidly it converges to a globally optimal assignment and topology -- is governed by the connectivity and properties of the communication network established between the material particles (see Fig.~\ref{fig:F1}c). The autonomous metamaterial is, structurally, a \textit{distributed networked system}~\rev{\citep{mesbahi2010,amirkhani2022}}: the particles are \textit{nodes}, and the wireless communication links are \textit{edges} of the network. At any instant during deformation, this communication network of the autonomous metamaterial is represented by a graph $\mathcal{G}=(\mathcal{V},\mathcal{E})$, where the vertex set $\mathcal{V}=\{1, 2,...,n\}$ indicates $n$ nodes (particles) in the graph, and the edge set $\mathcal{E}$ contains an edge $(i,k)$ if and only if particles $i$ and $k$ lie within each other's communication zone, i.e., $\mathcal{E}=\{(i,k):\|\textbf{X}_{\pi(i)}-\textbf{X}_{\pi(k)}\|\leq r_c ,i\ne k\}$~\citep{penrose2003,godsil2001}. The convergence of Algorithm~II to a globally optimal assignment, and consequently the ability of the metamaterial to coordinate a topology change, depends critically on the connectivity of $\mathcal{G}$: connectivity is precisely the condition under which every particle's cost information reaches every other particle through the neighbor communication chain. 

The structure of the graph $\mathcal{G}$ is encoded in the adjacency matrix $A$, with $A_{ik} = 1$ if $(i,k)\in \mathcal{E}$ and $A_{ik} = 0$ otherwise, and in the degree matrix $D=\mathrm{diag}(d_1,..., d_n)$, where $d_i=\sum_{k\in\mathcal{N}_i} A_{ik}$ is the number of communication neighbors of particle $i$~\citep{godsil2001,mesbahi2010}. The network connectivity is characterized based on the graph Laplacian matrix~\rev{\citep{fiedler1973,olfatisaber2007,mesbahi2010,amirkhani2022,olfatisaber2004}}: 
\begin{equation}
L=D-A.
\label{eq:4}
\end{equation}
The second-smallest eigenvalue $\lambda_{2}$ of the Laplacian $L$ is the algebraic connectivity, and it is used here to examine the coordination performance of the autonomous metamaterial: when $\lambda_{2}>0$, the communication network is connected and permits information to propagate throughout the metamaterial, thereby enabling globally coordinated topology evolution; when $\lambda_{2}=0$, the network is disconnected and fragments into independent sub-networks that cannot collectively optimize the material topology. In the present work, the decentralized coordination process is executed only for autonomous metamaterials with connected communication networks ($\lambda_2>0$), and the process is terminated when the metamaterial fragments into disconnected sub-networks. 

More quantitatively, $\lambda_{2}$ sets the rate of information propagation and consensus among the material particles: for a linear consensus process on $\mathcal{G}$ the disagreement decays as $\exp(-\lambda_{2}t)$ \citep{olfatisaber2007}, so that the number of coordination iterations required for the decentralized assignment to converge scales inversely with $\lambda_{2}$. Fig.~\ref{fig:F1}c demonstrates this behavior for the autonomous metamaterial, where the algebraic connectivity $\lambda_2$ increases with the communication radius $r_{c}$, and the number of iterations $t$ required for convergence to the optimal topology decreases correspondingly \rev{($\lambda_2$ is evaluated after every load step, as described in~\ref{app:A2})}. 

Before moving forward, we make two remarks here. First, the decentralized coordination process described here is \textit{stochastic} at the particle level. In any asynchronous decentralized assignment process, multiple optimal assignments of equal total cost commonly exist~\rev{\citep{ismail2017,dey2010,huang2018}}. Therefore, under identical loading conditions, the decentralized process may assign the same particle to a different lattice site when the process is repeated. The framework proposed here does not eliminate this particle-level stochasticity, and does not attempt to. However, at the level of the emergent topology, the process is \textit{deterministic} -- and it is the topology $\mathcal{S}(\pi)=\text{image}(\pi)$, not the particle assignment $\pi$, that governs the emergent constitutive behavior. Because the material particles are identical and the underlying lattice is fixed, any permutation of identical particles among the same set of occupied sites produces the same topology, the same elastic-bond structure, and the same effective stiffness. If the particles are not identical, or if they are not constrained to a fixed lattice, the process could be non-deterministic in general. This defines the specific mechanism by which deterministic constitutive behavior emerges from an inherently stochastic decentralized process in the proposed autonomous metamaterial.

Second, while this deterministic mechanism ensures that different assignments of the particles over the same topology yield the same constitutive behavior, it does not by itself guarantee that the decentralized coordination process converges to the same topology when the process is repeated. Two further conditions are required: (a) the cost function along with the assignment problem must select a unique optimal topology; and (b) the communication network of the autonomous metamaterial must be sufficiently connected for the decentralized algorithm to reach that optimum. Thus, the assignment problem governs whether a unique optimal topology exists, while the connectivity governs whether it is reachable. 

\section{Deterministic Constitutive Behavior}
\label{sec:3}
Having established how the topology of the autonomous metamaterial evolves as a sequence of reassignment events executed through decentralized coordination among the material particles, we here demonstrate that such topology evolution is accompanied by spontaneous stiffness modulation, enabling the material to change its property landscape between achiral and chiral states. We further establish how this autonomous evolution can give rise to nonreciprocal elasticity, where the autonomous metamaterial can modulate its elastic properties depending on the direction of the applied load/deformation~\rev{\citep{shaat2020d,shaat2020a,shaat2023}}. An effective constitutive model is therefore developed to directly relate the evolving topology to the material moduli and emergent material symmetry, characterizing these capabilities as functions of loading history and the underlying decentralized intelligence of the metamaterial.

\subsection{Constitutive Model}
At a fixed topology, the autonomous metamaterial is a 2D linear elastic solid, and its homogenized response is governed by the classical Hooke's law: 
\begin{equation}
\sigma_{ij}=C_{ijkl} \epsilon_{kl}, 
\label{eq:5}
\end{equation}
where $\sigma_{ij}(=\sigma_{ji})$ is the Cauchy stress tensor, $\epsilon_{kl}(=\epsilon_{lk})$ is the strain tensor, and $C_{ijkl}$ is the fourth-order elasticity tensor. Autonomy does not modify this constitutive law, e.g., by introducing additional kinematic or stress variables, but rather it enables spontaneous modulation of the elastic constants of $C_{ijkl}$ through the fact that the topology that generates this law is selected by the material itself. 

To explore how topology evolution changes the material properties landscape, the constitutive relations in Eq.~(\ref{eq:5}) are rewritten in terms of three volumetric deformation modes -- dilation $\gamma_1$, shear~1 $\gamma_2$ and shear~2 $\gamma_3$~\rev{\citep{shaat2025b,shaat2026b}}:
\begin{equation}
\gamma_{1}=\epsilon_{xx}+\epsilon_{yy},
\qquad
\gamma_{2}=\epsilon_{xx}-\epsilon_{yy},
\qquad
\gamma_{3}=2\epsilon_{xy},
\label{eq:6}
\end{equation}
so that
\begin{equation}
\begin{bmatrix}	
\sigma_1 \\
\sigma_2\\
\sigma_3
\end{bmatrix}
=
\begin{bmatrix}
B & \eta & \xi_1\\
\eta & 2\mu+\Gamma & \xi_2\\
\xi_1 & \xi_2 & 2\mu-\Gamma
\end{bmatrix}
\begin{bmatrix}
\gamma_{1}\\
\gamma_{2}\\
\gamma_{3}
\end{bmatrix}
,
\qquad
\sigma_{\alpha}=C_{\alpha\beta}\,\gamma_{\beta},
\label{eq:7}
\end{equation}
where the components of the  stress vector $\sigma_\alpha$ take the form:
\begin{equation}
\sigma_{1}=\sigma_{xx}+\sigma_{yy}, \qquad
\sigma_{2}=\sigma_{xx}-\sigma_{yy}, \qquad
\sigma_{3}=2\sigma_{xy}.
\label{eq:8}
\end{equation}
and $C_{\alpha\beta}$ ($\forall (\alpha,\beta)\in\{1,2,3\}$) is the effective stiffness matrix of the autonomous metamaterial.

In the stiffness matrix $C_{\alpha\beta}$ of Eq.~\eqref{eq:7}, $B$ and $\mu$ are the bulk and shear moduli, $\eta$ and $\Gamma$ encode anisotropy, and $\xi_{1}$ and $\xi_{2}$ encode chirality as well as anisotropy~\rev{\citep{shaat2025a,shaat2025b,shaat2026b}}. For topologies of $C_3$ or $C_6$ symmetry, the stiffness matrix is isotropic and achiral with $\eta=\Gamma=\xi_1=\xi_2=0$. When the topology evolves to $C_4$ symmetry, the stiffness matrix shows anisotropy through a non-zero $\Gamma$, and both anisotropy and chirality when $\xi_2$ is non-zero. When the topology changes to $C_1$ or $C_2$ symmetry, the matrix takes the general form in Eq.~\eqref{eq:7}, reflecting a chiral, anisotropic material. The matrix $C_{\alpha\beta}$ therefore carries the complete information on the material's symmetry class and on its elastic moduli, and it is this matrix that will be shown in the subsequent sections to evolve autonomously.

The procedure used to determine the components of the evolving stiffness matrix $C_{\alpha\beta}$ of the autonomous metamaterial at each topology state is detailed in~\ref{app:A4} \rev{and~\ref{app:A5}}.

\subsection{Autonomous Stiffness Modulation}

\begin{figure}[t]
\centering
\includegraphics[width=\textwidth]{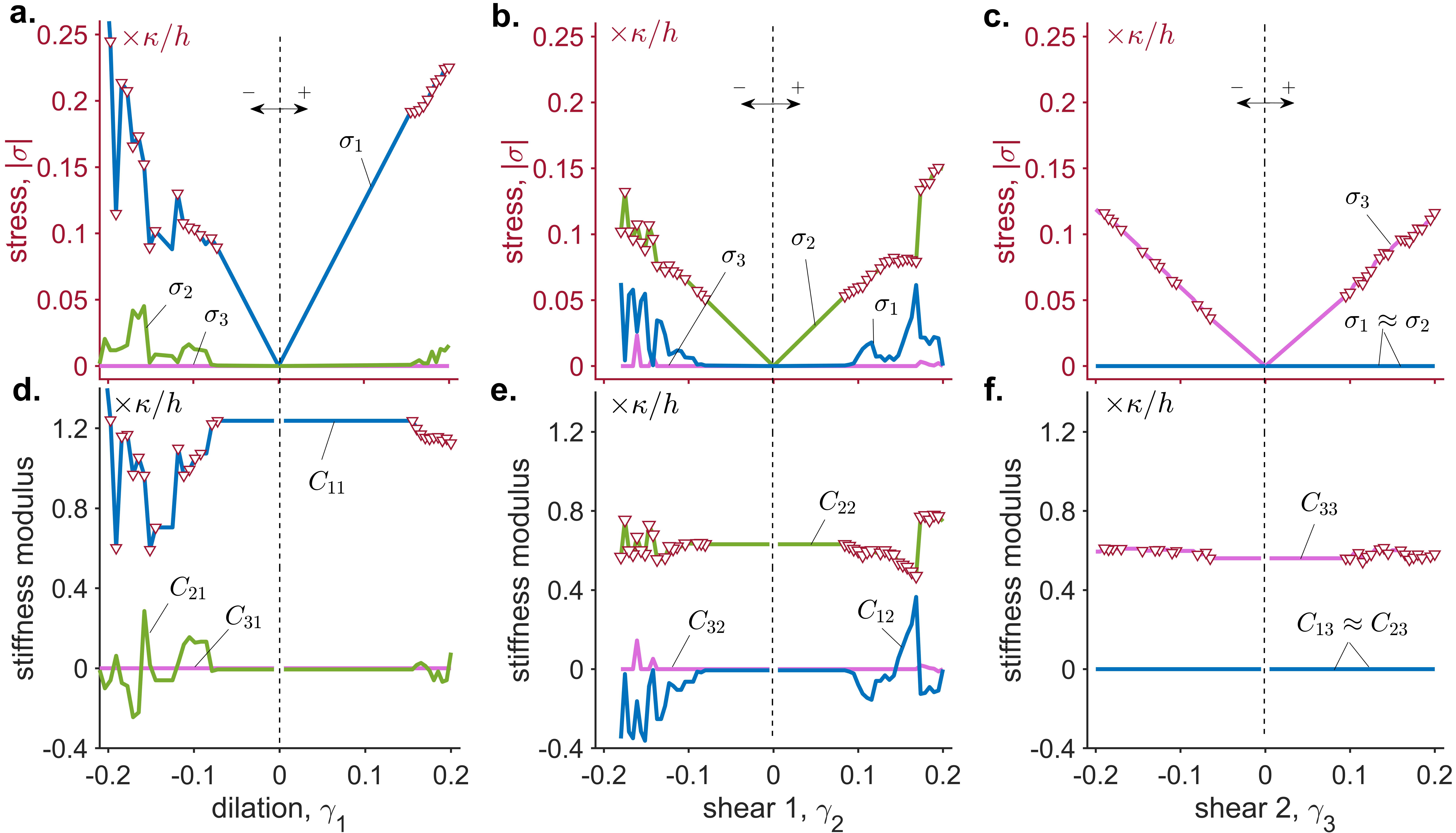}
\caption{\rev{\textbf{Autonomous stiffness modulation and nonreciprocity.} \textbf{a}--\textbf{c.} Magnitudes of the stress components $|\sigma_\alpha|$ versus the applied deformation for \textbf{a.} dilation $\gamma_1$, \textbf{b.} shear~1 $\gamma_2$ and \textbf{c.} shear~2 $\gamma_3$, each applied monotonically from the initial topology $\mathcal{S}_0$ in the negative ($-$) and positive ($+$) directions, demonstrating nonreciprocal elastic behavior. \textbf{d}--\textbf{f.} Corresponding components of the effective stiffness matrix $C_{\alpha\beta}$. Markers indicate topology-evolution events. The computation of stresses and stiffnesses are detailed in~\ref{app:A3}, \ref{app:A4} and \ref{app:A5}.}}
\label{fig:F2}
\end{figure}

Along a deformation path, the autonomous metamaterial may undergo a sequence of topology-evolution events $\mathcal{S}_{0}\to\mathcal{S}_{1}\to\cdots\to\mathcal{S}_{m}$, after each of which, the elastic bond network, and therefore the stiffness, changes (see Fig.~\ref{fig:F1}b). This is \emph{autonomous stiffness modulation}: a stiffness that is modulated along the loading path, not by an external controller or an applied stimulus, but by the material's own decentralized decisions. Figure~\ref{fig:F2} demonstrates this behavior for an autonomous metamaterial whose unit cell has the initial topology shown in Fig.~\ref{fig:F1}a, subjected separately to dilation $\gamma_1$, shear~1 $\gamma_2$ and shear~2 $\gamma_3$ in both the positive and negative directions \rev{(the boundary conditions and deformation paths are given in~\ref{app:A3})}. Initially, the stress response $\sigma_\alpha$ varies linearly with the deformation $\gamma_\beta$ (Figs.~\ref{fig:F2}a-c), such that the metamaterial exhibits a constant stiffness $C_{\alpha\beta}$, $\forall (\alpha,\beta)\in \{1,2,3\}$ (Figs.~\ref{fig:F2}d-f). As deformation increases, the particles eventually assign themselves to new lattice sites, changing topology, and with it the stiffness moduli $C_{\alpha\beta}$ change. The stiffness therefore remains constant between events and changes only when the metamaterial undergoes a topology-evolution event (Figs.~\ref{fig:F2}d-f). This autonomous stiffness modulation \rev{does not result from a modification of the material's constitutive law; it is driven solely by the material's internal intelligence and its own decisions}. 

A distinctive behavior of the autonomous metamaterial is its ability to evolve not only its elastic moduli with each topology-change event but also its behavior landscape, switching it between achiral and chiral material responses. Before any topology-change event is triggered, the stiffness matrix is diagonal, with $C_{11}=B$, $C_{22}=2\mu+\Gamma$ and $C_{33}=2\mu-\Gamma$ \rev{as the only} non-zero moduli (Figs.~\ref{fig:F2}d-f), so that the autonomous metamaterial is initially anisotropic but achiral. As the metamaterial deforms and its topology evolves, this landscape changes, where the diagonal moduli change and new material moduli emerge, depending on the deformation path. For instance, under the dilation $\gamma_{1}$ path shown in Figs.~\ref{fig:F2}a,d, the anisotropy modulus $C_{21}=\eta$ becomes non-zero while the chiral modulus $C_{31}=\xi_{1}$ remains zero, as confirmed by the vanishing stress $\sigma_{3}$: the metamaterial develops a new anisotropic coupling between dilation and shear~1 but remains achiral. Under shear~1 $\gamma_{2}$ (Figs.~\ref{fig:F2}b,e), the anisotropy modulus $C_{12}=\eta$ likewise becomes non-zero, and at isolated events the chiral modulus $C_{32}=\xi_{2}$ departs from zero as well, so that the metamaterial becomes chiral and anisotropic. Under shear~2 $\gamma_{3}$ (Figs.~\ref{fig:F2}c,f), by contrast, numerous topology-evolution events occur, yet the chiral moduli $C_{13}=\xi_{1}$ and $C_{23}=\xi_{2}$ remain zero and $C_{33}$ changes only slightly, so that the character of the material response is preserved and the metamaterial remains achiral. The autonomous metamaterial can therefore transition between achiral and chiral, and between isotropic and anisotropic, responses along a single loading path, as its topology evolves.

We emphasize that the stress variations with the deformations appear in Figs.~\ref{fig:F2}a-c as changes of slope rather than as a nonlinear constitutive response. The metamaterial is linear elastic between events, exhibiting linear stress--deformation relations. The apparent nonlinearity of the stress--deformation curves \rev{results entirely from} the accumulated effect of topology evolution, driven here by the material's intelligence \rev{rather than by} a nonlinear constitutive response. 

\subsection{Direction-dependent Topology Evolution and Nonreciprocity}
Mechanical nonreciprocity requires breaking one of the assumptions underlying the Onsager-Casimir reciprocity relations~\rev{\citep{casimir1945,asadchy2020,nassar2020}}. Established routes to nonreciprocity include (i) \textit{breaking time-reversal symmetry} through a momentum bias produced by active feedback controls~\rev{\citep{brandenbourger2019}}, gyroscopic effects~\rev{\citep{shaat2024a,nash2015}} or spatiotemporal modulation of stiffness or inertia~\rev{\citep{trainiti2016,nassar2017}}; (ii) \textit{breaking parity (spatial) inversion symmetry} through geometric or contact nonlinearity in asymmetric structures~\rev{\citep{coulais2017,shaat2020a,shaat2020d,xu2026,shobeiri2025}}; and (iii) \textit{breaking thermodynamic equilibrium} through asymmetric elasticity tensors that characterize net work exchange over closed deformation cycles, as in odd elasticity~\rev{\citep{scheibner2020b}} and chiral nonreciprocal elasticity~\rev{\citep{shaat2023}}. 

The autonomous metamaterial developed here exhibits mechanical nonreciprocity that falls within the third route but arises through a fundamentally novel mechanism whose origin lies in the \emph{direction-dependence of its topology evolution}. When the metamaterial is deformed along a given axis, the displacement field generated and the decentralized coordination process drive particle reassignments, which produce a specific sequence of topology change events $\mathcal{S}_{0}\to\mathcal{S}^{+}_{1}\to \mathcal{S}^{+}_{2}\to\dots\to\mathcal{S}^{+}_{m}$. When the deformation is reversed along the same axis -- for example, switching from tension $+\gamma_{1}$ to compression $-\gamma_{1}$ -- the displacement field reverses, changing the cost landscape and driving the particles toward different lattice sites, thereby producing a different sequence of topology change events $\mathcal{S}_{0}\to\mathcal{S}^{-}_{1}\to\mathcal{S}^{-}_{2}\to\dots\to\mathcal{S}^{-}_{m'}$. Thus, the decentralized coordination process follows different sequences of topology-evolution events depending not only on the deformation mode but also \rev{on its direction}, producing a nonreciprocal elastic response, where the effective stiffness becomes direction-dependent:
\begin{equation}
C_{\alpha\beta}^{+}\neq C_{\alpha\beta}^{-},
\label{eq:9}
\end{equation}
where $C_{\alpha\beta}^+$ and $C_{\alpha\beta}^-$ are the effective stiffness values reached under positive $\gamma_\beta^+$ and negative $\gamma_\beta^-$ deformations, respectively (see Fig.~\ref{fig:F2}). This is precisely the type of nonreciprocity developed in our prior work~\rev{\citep{shaat2023,shaat2020d}}, in which the components of the stiffness matrix acquire different values under opposite deformation modes:
\begin{equation}
C_{\alpha\beta}^\pm=\begin{bmatrix}
B^\pm & \eta^\pm & \xi_1^\pm\\
\eta^\pm & 2\mu^\pm+\Gamma^\pm & \xi_2^\pm\\
\xi_1^\pm & \xi_2^\pm & 2\mu^\pm-\Gamma^\pm
\end{bmatrix} .
\label{eq:10}
\end{equation}

This nonreciprocal behavior is demonstrated in Fig.~\ref{fig:F2} where the stress--deformation curves, and therefore the associated stiffness moduli $C_{\alpha\beta}$, obtained under the different deformation modes are not identical when inverting the applied deformation along the same direction. At small deformations, and when no event has yet been triggered, the autonomous metamaterial exhibits reciprocal behavior, where identical stress--deformation curves and the same stiffness values are obtained under the two directions. When the metamaterial undergoes topology-evolution events in which the sequences of topologies produced under $+\gamma_{\beta}$ and $-\gamma_{\beta}$ are distinct, the emergent elastic behavior becomes nonreciprocal. It is worth noting that this nonreciprocity does not \rev{arise from} a momentum source, an applied non-conservative body force, or an invoked geometric nonlinearity, but rather it is generated by the autonomous decision process that selects the sequence of topologies that emerge over the deformation path. 

\section{Mechanical Activity}
\label{sec:4}

\begin{figure}[t]
\centering
\includegraphics[width=\textwidth]{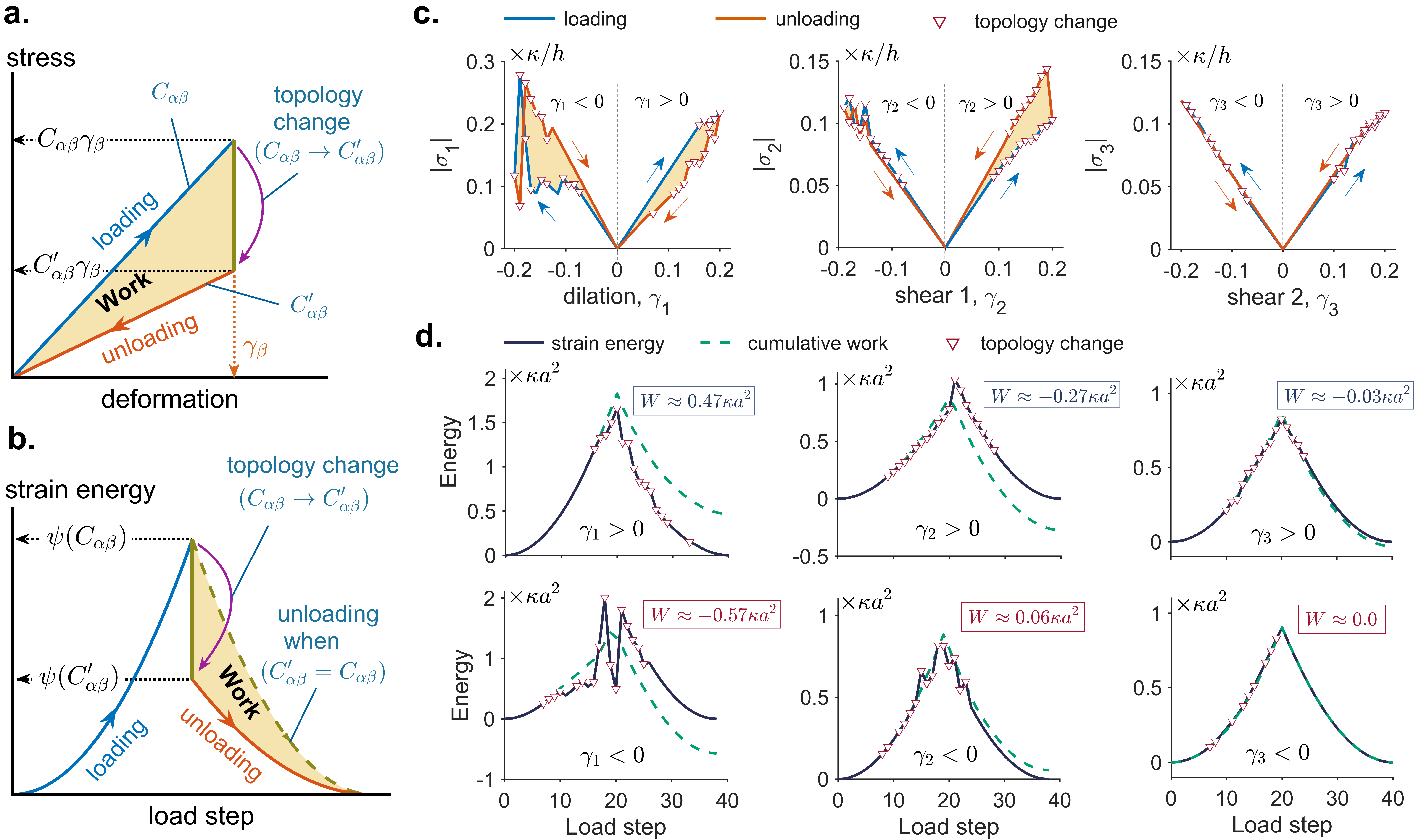}
\caption{\rev{\textbf{Mechanical activity.} \textbf{a}--\textbf{b.} Illustration of the mechanism of mechanical activity: \textbf{a.} stress versus deformation and \textbf{b.} strain energy $\psi$ versus load step over a closed quasistatic deformation cycle in which a topology-evolution event at $\gamma_\beta$ changes the stiffness from $C_{\alpha\beta}$ to $C^\prime_{\alpha\beta}$; the dashed curve in \textbf{b} shows unloading without a topology change. \textbf{c}--\textbf{d.} Mechanical activity of the autonomous metamaterial: \textbf{c.} stress magnitudes $|\sigma_\alpha|$ over closed cycles of dilation $\gamma_1$, shear~1 $\gamma_2$ and shear~2 $\gamma_3$, each applied in the negative and positive directions; \textbf{d.} corresponding strain energy $\psi$ and cumulative work versus load step for positive (top row) and negative (bottom row) deformations, with the net work $W$ indicated in each panel. Shaded areas indicate the net work exchanged over each cycle. The deformation cycles and the computation of strain energy and cumulative work are detailed in~\ref{app:A3} and~\ref{app:A6}.}}
\label{fig:F3}
\end{figure}

\subsection{Energy Balance}
The energy balance of the autonomous metamaterial when it undergoes reversible, isothermal processes can be defined based on the first law and second law of thermodynamics as~\rev{\citep{shaat2023}}
\begin{equation}
\mathrm{\delta} W = \mathrm{d} \psi , 
\label{eq:11}
\end{equation}
where $\mathrm{d}\psi$ is the net change in the Helmholtz free energy of the autonomous metamaterial, while $\mathrm{\delta}W$ is the net mechanical work transferred through its boundary. 

The free energy $\psi$ in Eq.~\eqref{eq:11} is a state function, and thus the integral of $\mathrm{d}\psi$ over a deformation cycle $\Omega$ depends only on the initial and final states: 
\begin{equation}
\oint_{\partial \Omega}{\mathrm{d}\psi}=\psi_{\mathrm{final}}-\psi_{\mathrm{initial}}.
\label{eq:12}
\end{equation}
The work $\mathrm{\delta} W$ however is not a state variable, as its integral over a deformation cycle $\oint_\Omega{\mathrm{\delta} W}$ depends on the path taken through the cycle. This distinction establishes the condition at which the autonomous metamaterial exhibits mechanical activity: 
\begin{equation}
W=\oint_{\partial \Omega}{\mathrm{\delta}} W-\oint_{\partial \Omega}{\mathrm{d}}\psi \ne 0,
\label{eq:13}
\end{equation}
that is, the work used to drive the metamaterial through the deformation cycle $\oint_{\partial \Omega}{\mathrm{\delta}} W$ is not equal \rev{to} the net change in its free energy through the cycle $\oint_{\partial \Omega}{\mathrm{d}}\psi$~\rev{\citep{shaat2023}}. The difference between $\oint_{\partial \Omega}{\mathrm{\delta}} W$ and $\oint_{\partial \Omega}{\mathrm{d}}\psi$ is a cumulative, mechanical work $W$ exchanged between the metamaterial and its surroundings over the deformation cycle. A non-zero mechanical work $W\ne 0$ is a signature of mechanical activity, where $W>0$ denotes work done \emph{on} the metamaterial by the surroundings and $W<0$ \rev{denotes work delivered} \emph{to} the surroundings across the metamaterial boundary. \rev{Only when $W=0$ is the energy conserved}, and the autonomous metamaterial does not exhibit any mechanical activity. 

Adopting the constitutive model in Eqs.~\eqref{eq:6}--\eqref{eq:8}, the energy balance in Eq.~\eqref{eq:11} gives the work needed to deform the metamaterial through a quasistatic deformation path from zero to $\gamma_\alpha$ as (see~\rev{\ref{app:A6}})
\begin{equation}
\int_0^{\gamma_\alpha}{\mathrm{\delta}W} = \frac{1}{4}\int_V{{C_{\alpha\beta}\gamma_\beta \gamma_\alpha}} \mathrm{d}V, 
\label{eq:14}
\end{equation}
where $V$ is the volume of the autonomous metamaterial. 

\subsection{Mechanism of Mechanical Activity and Mechanical Work}
Recent advances in active mechanical systems have shown that active elastic solids can exchange net mechanical work with their surroundings. \textit{Odd elasticity} -- established in~\rev{\citet{scheibner2020b}} and experimentally realized in~\rev{\citet{veenstra2025,chen2021}} -- achieves this through non-conservative microscopic interactions that produce an asymmetric stiffness matrix that has no major symmetry $C_{\alpha\beta}\ne C_{\beta\alpha}$. \textit{Chiral nonreciprocal elasticity} -- developed in~\rev{\citet{shaat2023}} and implemented in~\rev{\citet{montazeri2025,yuan2025,shaat2026c}} -- enables mechanical activity through the combination of chirality with nonreciprocity, yielding an asymmetric elastic response that deviates during loading and unloading paths. 

The autonomous metamaterial realizes mechanical activity through a fundamentally different mechanism: \textit{autonomous stiffness modulation driven by topology evolution}. Although the metamaterial remains elastic at every instant and therefore satisfies the energy balance in Eq.~\eqref{eq:11}, it can undergo topology-change events that alter its stiffness, resulting in a non-zero mechanical work $W\ne 0$ exchanged with the surroundings over a deformation cycle. 

This mechanism is illustrated in Figs.~\ref{fig:F3}a,b for an autonomous metamaterial undergoing a closed quasistatic deformation cycle $\Omega$. Because the deformation cycle is closed, the metamaterial recovers its undeformed state, and its \rev{free energy returns} to zero at the end of the cycle, i.e., $\psi_{\mathrm{final}}=\psi_{\mathrm{initial}}=0$, and therefore -- according to Eq.~\eqref{eq:12} -- the net change in its free energy across the cycle is zero, i.e., $\oint_{\partial \Omega}{\mathrm{d}\psi}=0$. The total work driving the metamaterial through the cycle $\oint_{\partial\Omega}{\mathrm{\delta}W}$, however, need not \rev{vanish}. The deformation is first increased from zero to $\gamma_{\alpha}$ along a loading path, over which the work required to drive the metamaterial follows from Eq.~\eqref{eq:14} as $\int_{0}^{\gamma_{\alpha}}\delta W=\tfrac{1}{4}\int_V {C_{\alpha\beta}\gamma_{\beta}\gamma_{\alpha}}\mathrm{d}V$, $\forall(\alpha,\beta)\in\{1,2,3\}$ (Fig.~\ref{fig:F3}b). The deformation is then reversed from $\gamma_{\alpha}$ back to zero in an unloading path. In the absence of autonomy, the initial topology $\mathcal{S}_{0}$ and hence the stiffness would persist throughout the cycle, and when the metamaterial is unloaded, it would return exactly the work applied on loading. However, because the metamaterial is autonomous, one or more topology-evolution events occur, changing the stiffness from $C_{\alpha\beta}$ during loading to $C^\prime_{\alpha\beta}$ during unloading, so that the metamaterial returns only $\int_{\gamma_{\alpha}}^{0}\delta W=-\tfrac{1}{4}\int_V{C^\prime_{\alpha\beta}\gamma_{\beta}\gamma_{\alpha}}\mathrm{d}V$ (Fig.~\ref{fig:F3}b). This difference gives a non-zero net work performed over the closed deformation cycle, i.e., $\oint_{\partial\Omega}{\mathrm{\delta}W}=\int_{0}^{\gamma_\alpha}{\mathrm{\delta}W}+\int_{\gamma_\alpha}^{0}{\mathrm{\delta}W}\ne 0$, producing -- according to Eqs.~\eqref{eq:13} and \eqref{eq:14} -- a cumulative work exchange with the surroundings of:
\begin{equation}
W=\frac{1}{4}\int_V{\bigl(C_{\alpha\beta}-C^\prime_{\alpha\beta}\bigr) \gamma_{\beta}\gamma_{\alpha}}\mathrm{d}V.
\label{eq:15}
\end{equation}

This mechanical activity is further demonstrated in Figs.~\ref{fig:F3}c,d, which show how topology evolution drives autonomous stiffness modulation and enables the metamaterial to exchange non-zero mechanical work with its surroundings over closed quasistatic deformation cycles. Figure~\ref{fig:F3}c shows the stress--deformation responses of the autonomous metamaterial over closed cycles of dilation $\gamma_1$, shear~1 $\gamma_2$ and shear~2 $\gamma_3$, each applied in the positive and negative directions, with markers identifying the topology-change events \rev{(the deformation cycles are defined in~\ref{app:A3})}. These events produce changes in the stress response and, more importantly, modulate the stiffness, so that the resulting loading and unloading paths do not generally coincide, indicating a nonzero net work exchange over the deformation cycle. This mechanical activity is quantified in Fig.~\ref{fig:F3}d, which compares the evolution of the strain energy $\psi$ with the cumulative work for each deformation cycle\rev{, both computed as described in~\ref{app:A6}}. At the end of each cycle, the strain energy returns to zero as the metamaterial recovers its undeformed state, whereas the cumulative work settles at a non-zero value, corresponding to the net work $W$ exchanged with the surroundings. 

The results in Figs.~\ref{fig:F3}c,d also demonstrate that the same autonomous metamaterial acts either as an energy sink or as an energy source, depending on the deformation mode and its direction. Which role it
plays is set by the free-energy changes at the topology-evolution events (see Fig.~\ref{fig:F3}d). \rev{For instance}, when the autonomous metamaterial is deformed under tensile dilation \rev{to} $\gamma_1=+0.2$ \rev{and back}, a positive net work of $W\approx+0.47\kappa a^2$ \rev{(energies are normalized by $\kappa a^2$; see~\ref{app:A})} is obtained, indicating that the surroundings recover less work during unloading than they supplied during loading, and the metamaterial acts as an energy sink. The metamaterial does not store this energy, since its free energy returns to zero at the end of the cycle (Fig.~\ref{fig:F3}d). The energy is instead released at the topology-evolution events, at which particles detach from strained bonds and the bond network relaxes at fixed deformation. Conversely, when the metamaterial is compressed to $\gamma_{1}=-0.2$ \rev{and back}, the strain energy generally rises at the topology-evolution events, and the metamaterial delivers a net work of $0.57\kappa a^2$ ($W\approx-0.57\kappa a^2$) to its surroundings. This energy is generated as the particles establish new bonds within the strained elastic lattice, and it is in this regime that the metamaterial acts as a genuine energy source. Shear~1 shows the same duality with the roles reversed: deformed to $\gamma_{2}=+0.2$ and unloaded, the metamaterial delivers $0.27\kappa a^2$ to the surroundings ($W\approx-0.27\kappa a^2$), whereas deformed to $\gamma_{2}=-0.2$ and unloaded, it dissipates $W\approx0.06\kappa a^2$. This direction dependence is a manifestation of the nonreciprocity
established in Fig.~\ref{fig:F2}.

The magnitude of the work $W$ reflects how much the topology-evolution events change the stiffness of the metamaterial. This activity is not a direct manifestation of topology evolution, as mechanical work emerges only insofar as the autonomous topology evolution modulates the stiffness. For example, under shear~2, numerous events occur in both directions, yet the net work is only $W\approx-0.03\kappa a^2$ for $\gamma_{3}=+0.2$ and $W\approx0$ for $\gamma_{3}=-0.2$. The metamaterial is nearly conservative under shear~2 because the topology-evolution events barely change the stiffness (Figs.~\ref{fig:F2}c,f), and therefore the free-energy changes are correspondingly small, and the strain energy and the cumulative work remain nearly coincident throughout the cycle (Fig.~\ref{fig:F3}d).

We emphasize that the autonomous metamaterial is linear elastic within every topology state, contains no
actuators acting on its deformation, and follows no prescribed feedback law, and thus it always \rev{satisfies} the energy balance in Eq.~\eqref{eq:11}. Therefore, its mechanical activity and the work it exchanges with the surroundings arise solely from the decisions by which it reorganizes its own topology. Autonomous stiffness modulation is thus a route to mechanical activity distinct from existing approaches.

\section{Conclusions}
\label{sec:5}

On the one hand, solid materials have so far been passive recipients of decisions made elsewhere: decisions encoded at synthesis, delivered through external stimuli, or issued by a controller. On the other hand, decentralized systems capable of making their own decisions, such as robotic swarms, generally lack a constitutive description because their collective behavior is stochastic. This work bridges these two paradigms and demonstrates that a solid can be endowed with genuine decentralized autonomy, making its own decisions while retaining a deterministic constitutive law. The key insight is that determinism need not be imposed on the decision-making process itself; rather, it can emerge at the level of the autonomous material's collective topology. 

Establishing determinism makes the mechanics of autonomy accessible and reveals behaviors with no direct counterpart in existing material paradigms. The stiffness of the autonomous metamaterial is no longer a fixed material constant but a state variable modulated by the material itself. This autonomous stiffness modulation enables the material to dynamically activate anisotropic and chiral moduli along a loading path, even though its constitutive response remains linear elastic within each topology state. Moreover, because the sequence of topology-evolution events depends on the direction of deformation, the resulting autonomous stiffness modulation produces nonreciprocity of a new origin, one that requires neither a momentum bias, nor geometric nonlinearity, nor prescribed asymmetric moduli. The dependence of topology evolution on loading history also establishes a new route to adaptive mechanical activity: the same material exchanges net work with its surroundings over closed deformation cycles, acting as an energy sink or an energy source depending on the deformation mode and its direction.

These findings position autonomy as a design principle for solid materials, in which material laws and constitutive behavior emerge from local, decentralized rules rather than being fixed at synthesis or prescribed externally. In this setting, local interaction and decision-making rules become design variables that can be exploited to develop materials that self-adapt, self-heal or process information through their own reorganization. Future physical realization of such materials, for instance with miniature mobile robots~\citep{rubenstein2014b}, could bring decentralized intelligence demonstrated in robotic collectives into the realm of synthetic materials and constitutive mechanics.

\section*{Declaration of competing interest}
The author declares no competing financial interests or personal relationships that could have appeared to influence the work reported in this paper.

\section*{Data availability}
Data will be made available on request. 

\section*{Acknowledgements}
The author gratefully acknowledges the support from St. Mary’s University through the StMU-SwRI Technology \& Applied Research program. 

\appendix

\section{Computational Implementation}
\label{app:A}

All simulations use a MATLAB implementation of the autonomous metamaterial model presented in Section~\ref{sec:2}. Results are reported in units of the bond stiffness $\kappa$ and the lattice spacing $a$. Forces, stresses (and stiffnesses) and energies are thus normalized by $\kappa a$, $\kappa/h$ and $\kappa a^2$, respectively, with $h$ the out-of-plane thickness of the 2D metamaterial. 

\subsection{Elastic Lattice and Initial Topology}
\label{app:A1}

The host lattice is triangular, with $N_x\times N_y=14\times 14$ sites, such that the position of the $j$th lattice site is defined by
\begin{equation}
\mathbf{X}_j=a\,\bigl(p+\tfrac12\chi_q,\ \tfrac{\sqrt3}{2}q\bigr),
\label{eq:A1}
\end{equation}
where $p=1,\dots,N_x$, $q=1,\dots,N_y$, with $\chi_q=1$ for odd $q$ and $0$ otherwise. The initial topology $\mathcal{S}_0$ of the metamaterial's unit cell contains vacant sites within radius $R=2.5a$, centered at $\bigl(\tfrac{a}{2}(N_x+1),\ \tfrac{\sqrt3}{4}N_ya\bigr)$, while the \rev{remaining} sites are occupied by particles (Fig.~\ref{fig:F1}a). The assignment $\pi$ is stored as a particle-indexed vector, such that particle $i$ \rev{occupies} site $\pi(i)$, and the topology is its image $\mathcal{S}(\pi)=\{\pi(1),\dots,\pi(n)\}$. An elastic bond is established between particles $i$ and $k$ if and only if their lattice sites are neighbors, $\|\mathbf{X}_{\pi(i)}-\mathbf{X}_{\pi(k)}\|\rev{\le r_n}$, with the neighbor radius $r_n=a$ (a $1.05a$ bond cut-off is used). Each bond is a linear spring of stiffness $\kappa$, \rev{and each of its two nodes has} $u_x$ and $u_y$ as \rev{its} two nodal degrees of freedom. The global stiffness $\mathbf{K}$, assembled across the unit cell, is partitioned into free (f) and prescribed (c) degrees of freedom, and equilibrium is solved from
\begin{equation}
\mathbf{K}_{ff}\,\mathbf{u}_f=-\mathbf{K}_{fc}\,\bar{\mathbf{u}}_c .
\label{eq:A2}
\end{equation}
where $\textbf{u}$ is the nodal displacement vector, and $\bar{\textbf{u}}_c$ is a vector of nodal displacements prescribed at the unit cell boundary.  

Every topology is stress-free in its reference lattice. A reassigned particle is placed on its new lattice site, and its new bonds form at rest length. No residual stress is carried across a reassignment: the next load step applies the \emph{total} boundary displacement to the new elastic network. Boundary particles are those on the extreme abscissae and ordinates of the occupied set. The elastic domain has in-plane dimensions $L_x=\bigl(N_x-\tfrac{1}{2}\bigr)a$ and $L_y=\tfrac{\sqrt{3}}{2}\,\bigl(N_y-1\bigr)a$, measured between the outermost lattice sites, and volume $V=L_xL_y h$.

, \qquad 

\subsection{Decentralized Coordination and Particle Reassignment}
\label{app:A2}

The coordination between the material particles is performed such that in each load step $\tilde{n}$, the equilibrium in Eq.~\eqref{eq:A2} is solved on the topology from the previous step $\mathcal{S}_{\tilde{n}-1}$. Then, the deformed positions of the particles are determined such that the deformed position of particle $i$ is $\mathbf{x}_i=\mathbf{X}_{\pi_{\tilde{n}-1}(i)}+\mathbf{u}_i$, where $\mathbf{u}_i$ is its displacement vector, and $\mathbf{X}_{\pi_{\tilde{n}-1}(i)}$ is the position of its previously assigned lattice site. Algorithm~II \rev{is} then executed to return an assignment $\pi_{\tilde{n}}$, and the particles are positioned at their reassigned sites $\mathbf{X}_{\pi_{\tilde{n}}(i)}$. The elastic bonds are then rebuilt if the new assignment differs from the previous assignment $\pi_{\tilde{n}}\neq\pi_{\tilde{n}-1}$. The cost in Eq.~\eqref{eq:2} measures the \emph{deformed} particle position against a candidate lattice site. The sets $\mathcal{J}_i$ and $\mathcal{N}_i$ are formed for each particle $i$ from its deformed position. 

Algorithm~II is executed synchronously such that, at iteration $t$, every particle reads its neighbors' cost matrices from iteration $t-1$. For each neighbor $k$, particle $i$ (a) copies every entry $c^k_{rj}$, $r\neq i$, that is finite in $\mathbf{c}^k$; and (b) for every lattice site $j$ revealed by $k$ in $\mathbf{c}^k$ (a finite entry in any row) that it does not yet know, it computes its own entry $c^i_{ij}$. A particle never copies its own row. The iterations stop when each particle does not receive new information from its neighbors, implying that all $n$ particles \rev{have reached the same local matrix}, verified within $10^{-10}$ consensus tolerance. Entries still unknown at convergence (sites outside every communication zone) are given the penalty cost $10^9$. The assignment problem of Eq.~\eqref{eq:1} on the common matrix is solved with an exact linear-assignment solver (MATLAB \texttt{matchpairs}), which returns the same optimum as Algorithm~I. A topology-evolution event is recorded at step $\tilde{n}$ when $\pi_{\tilde{n}}\neq\pi_{\tilde{n}-1}$. The algebraic connectivity $\lambda_2$ is evaluated after every step from Eq.~\eqref{eq:4}, with $A_{ik}=1$ for $\|\mathbf{X}_{\pi(i)}-\mathbf{X}_{\pi(k)}\|\le r_c$, $i\neq k$ and $D=\mathrm{diag}(\sum_{k\in\mathcal{N}_1}{A_{1k}},\sum_{k\in\mathcal{N}_2}{A_{2k}},\dots,\sum_{k\in\mathcal{N}_n}{A_{nk}})$.

\subsection{Deformation Paths and Cycles}
\label{app:A3}

The results presented in Figs.~\ref{fig:F1}b,~\ref{fig:F2}a-f, and~\ref{fig:F3}c,d are obtained by deforming the autonomous metamaterial under dilation $\gamma_1$, shear~1 $\gamma_2$ and shear~2 $\gamma_3$. Each deformation mode is imposed by prescribing one displacement component on each edge of the unit cell, while the other component is set free:
\begin{equation}
\begin{aligned}
\gamma_1:&\quad u_x|_L=0,\ u_x|_R=\bar\varepsilon L_x,\ u_y|_B=0,\ u_y|_T=\bar\varepsilon L_y,\\
\gamma_2:&\quad u_x|_L=0,\ u_x|_R=\bar\varepsilon L_x,\ u_y|_B=0,\ u_y|_T=-\bar\varepsilon L_y,\\
\gamma_3:&\quad u_y|_L=0,\ u_y|_R=\bar\varepsilon L_x,\ u_x|_B=0,\ u_x|_T=\bar\varepsilon L_y,
\end{aligned}
\label{eq:A3}
\end{equation}
with $\bar\varepsilon=\lambda\hat\gamma/2$, so that the nominal amplitude of the deformation mode is $\gamma_\beta=\lambda\hat\gamma$, $\forall \beta\in\{1,2,3\}$. The deformation is applied incrementally over $\tilde{N}$ load steps, such that the load factor at step $\tilde{n}$ is $\lambda_{\tilde{n}}=\tilde{n}/\tilde{N}$ on a monotonic deformation path (Figs.~\ref{fig:F1}b and \ref{fig:F2}a-f), and $\lambda_{\tilde{n}}=2\tilde{n}/\tilde{N}$ ($\tilde{n}\le \tilde{N}/2$), $\lambda_{\tilde{n}}=2-2\tilde{n}/\tilde{N}$ ($\tilde{n}>\tilde{N}/2$) on a closed deformation cycle (Figs.~\ref{fig:F3}c,d), which returns exactly to $\gamma_\beta=0$ at the end of the cycle. The positive ($\hat\gamma>0$) and negative ($\hat\gamma<0$) deformations are performed in separate runs, each starting from the initial topology $\mathcal{S}_0$.

\subsection{Stress and Strain Calculations}
\label{app:A4}

To calculate the variations of the stress and stiffness through a deformation path or a deformation cycle, the boundary forces and mean displacements on the right (R) and top (T) edges of the metamaterial's unit cell are computed at each load step as
\begin{equation}
\begin{gathered}
F_{xx}=\textstyle\sum_R f_x,\quad F_{xy}=\sum_R f_y,\quad
F_{yy}=\sum_T f_y,\quad F_{yx}=\sum_T f_x, \\
\bar u_{xx}=\langle u_x\rangle_{(R)}, \quad \bar u_{yx}=\langle u_y\rangle_{(R)}, \quad \bar u_{yy}=\langle u_y\rangle_{(T)}, \quad \bar u_{xy}=\langle u_x\rangle_{(T)},
\end{gathered}
\label{eq:A4}
\end{equation}
where $\bar u_{ab}$ is averaged over the edge on which $F_{ba}$ is summed. In Eq.~\eqref{eq:A4}, $f_x$ and $f_y$ are the components of reaction forces computed from $\mathbf{f}=\mathbf{K}\mathbf{u}$. Strains and stresses follow as
\begin{equation}
\begin{gathered}
\epsilon_{xx}=\frac{\bar u_{xx}}{L_x},\quad
\epsilon_{yy}=\frac{\bar u_{yy}}{L_y},\quad
\epsilon_{xy}=\frac{\bar u_{xy}}{L_y}, \quad
\epsilon_{yx}=\frac{\bar u_{yx}}{L_x},\\
\sigma_{xx}=\frac{F_{xx}}{L_y h},\quad
\sigma_{yy}=\frac{F_{yy}}{L_x h},\quad
\sigma_{xy}=\frac{F_{xy}}{L_y h},\quad
\sigma_{yx}=\frac{F_{yx}}{L_x h},
\end{gathered}
\label{eq:A5}
\end{equation}
and the deformation modes $\gamma_1$, $\gamma_2$ and $\gamma_3$ (Eq.~\eqref{eq:6}), and the corresponding stress vector components $\sigma_1$, $\sigma_2$ and $\sigma_3$ (Eq.~\eqref{eq:8}) are
\begin{equation}
\begin{gathered}
\gamma_1=\epsilon_{xx}+\epsilon_{yy},\quad
\gamma_2=\epsilon_{xx}-\epsilon_{yy},\quad
\gamma_3=\epsilon_{yx}+\epsilon_{xy}, \\
\sigma_1=\sigma_{xx}+\sigma_{yy},\quad
\sigma_2=\sigma_{xx}-\sigma_{yy},\quad
\sigma_3=\sigma_{xy}+\sigma_{yx}, 
\end{gathered}
\label{eq:A6}
\end{equation}

\subsection{Stiffness Calculations}
\label{app:A5}

The effective stiffness of the autonomous metamaterial at every topology state in the evolution sequence is determined by computing the slope of the stress-deformation curves at each load step. Specifically, the unit cell is subjected, in turn, to each of the three independent deformation modes $\gamma_{\beta}$ (with $\beta=1,2,3$). The resulting boundary forces are computed using Eq.~\eqref{eq:A4}, from which the corresponding stress components $\sigma_{\alpha}$ are obtained from Eq.~\eqref{eq:A5}, and the components of the stiffness matrix follow from
\begin{equation}
C_{\alpha\beta}
=\bigl[\sigma_{\alpha}^{(1)},\sigma_{\alpha}^{(2)},\sigma_{\alpha}^{(3)}\bigr]
\bigl[\gamma_{\beta}^{(1)},\gamma_{\beta}^{(2)},\gamma_{\beta}^{(3)}\bigr]^{-1},
\label{eq:A7}
\end{equation}
where the superscript denotes the applied load case.

\subsection{Strain Energy and Cumulative Work}
\label{app:A6}

The strain energy $\psi$ at each load step is computed as follows: 
\begin{equation}
\psi=\frac12\,\mathbf{u}^{\mathrm{T}}\mathbf{K}\,\mathbf{u}.
\label{eq:A8}
\end{equation}
where $\textbf{K}$ is the global stiffness matrix of the elastic unit cell, and $\textbf{u}$ is the displacement vector. \rev{These} strain energy calculations are verified based on the stress calculations from Eq.~\eqref{eq:A5}: 
\begin{equation}
\psi=\frac{V}{4} \left(\sigma_1\gamma_1+\sigma_2\gamma_2+\sigma_3\gamma_3 \right),
\label{eq:A9}
\end{equation}
which also verifies Eq.~\eqref{eq:14}. 

The \rev{cumulative} work is computed at each load step $\tilde{n}$ by accumulating the mechanical work 
based on the trapezoidal rule. For an applied deformation mode $\gamma_\alpha$, the boundary forces and mean displacements are obtained from Eq.~\eqref{eq:A4}, and subsequently the \rev{cumulative} work at the $\tilde{n}$th step is computed as
\begin{equation}
\begin{gathered}
W_{\tilde{n}}=\displaystyle\sum_{m=1}^{\tilde{n}} \bigl[ \frac{1}{2}\bigl(F_{xx}^{m}+F_{xx}^{m-1}\bigr)\bigl(\bar u_{xx}^{m}-\bar u_{xx}^{m-1}\bigr)+     \frac{1}{2}\bigl(F_{yy}^{m}+F_{yy}^{m-1}\bigr)\bigl(\bar u_{yy}^{m}-\bar u_{yy}^{m-1}\bigr) \bigr. \\
\bigl. +    \frac{1}{2}\bigl(F_{xy}^{m}+F_{xy}^{m-1}\bigr)\bigl(\bar u_{yx}^{m}-\bar u_{yx}^{m-1}\bigr)+    \frac{1}{2}\bigl(F_{yx}^{m}+F_{yx}^{m-1}\bigr)\bigl(\bar u_{xy}^{m}-\bar u_{xy}^{m-1}\bigr) \bigr], 
\end{gathered}
\label{eq:A10}
\end{equation}
 
When deformed through a closed cycle, the strain energy of the metamaterial returns to zero at the end of the cycle, i.e., $\psi_{\tilde{N}}=0$. The net work of Eq.~\eqref{eq:13} and in Fig.~\ref{fig:F3}d is therefore $W=W_{\tilde{N}}$ from Eq.~\eqref{eq:A10}.

\bibliographystyle{elsarticle-harv}

\begin{thebibliography}{112}
\expandafter\ifx\csname natexlab\endcsname\relax\def\natexlab#1{#1}\fi
\providecommand{\url}[1]{\texttt{#1}}
\providecommand{\href}[2]{#2}
\providecommand{\path}[1]{#1}
\providecommand{\DOIprefix}{doi:}
\providecommand{\ArXivprefix}{arXiv:}
\providecommand{\URLprefix}{URL: }
\providecommand{\Pubmedprefix}{pmid:}
\providecommand{\doi}[1]{\href{http://dx.doi.org/#1}{\path{#1}}}
\providecommand{\Pubmed}[1]{\href{pmid:#1}{\path{#1}}}
\providecommand{\bibinfo}[2]{#2}
\ifx\xfnm\relax \def\xfnm[#1]{\unskip,\space#1}\fi
\bibitem[{Adar et~al.(2004)Adar, Benenson, Linshiz, Rosner, Tishby and
  Shapiro}]{adar2004}
\bibinfo{author}{Adar, R.}, \bibinfo{author}{Benenson, Y.},
  \bibinfo{author}{Linshiz, G.}, \bibinfo{author}{Rosner, A.},
  \bibinfo{author}{Tishby, N.}, \bibinfo{author}{Shapiro, E.},
  \bibinfo{year}{2004}.
\newblock \bibinfo{title}{Stochastic computing with biomolecular automata}.
\newblock \bibinfo{journal}{Proceedings of the National Academy of Sciences}
  \bibinfo{volume}{101}, \bibinfo{pages}{9960--9965}.
\bibitem[{Amirkhani and Barshooi(2022)}]{amirkhani2022}
\bibinfo{author}{Amirkhani, A.}, \bibinfo{author}{Barshooi, A.H.},
  \bibinfo{year}{2022}.
\newblock \bibinfo{title}{Consensus in multi-agent systems: a review}.
\newblock \bibinfo{journal}{Artificial Intelligence Review}
  \bibinfo{volume}{55}, \bibinfo{pages}{3897--3935}.
\bibitem[{Anderson and Fernandez-Nieves(2022)}]{anderson2022}
\bibinfo{author}{Anderson, C.}, \bibinfo{author}{Fernandez-Nieves, A.},
  \bibinfo{year}{2022}.
\newblock \bibinfo{title}{Social interactions lead to motility-induced phase
  separation in fire ants}.
\newblock \bibinfo{journal}{Nature Communications} \bibinfo{volume}{13},
  \bibinfo{pages}{6710}.
\bibitem[{Anderson et~al.(2023)Anderson, Goldsztein and
  Fernandez-Nieves}]{anderson2023}
\bibinfo{author}{Anderson, C.}, \bibinfo{author}{Goldsztein, G.},
  \bibinfo{author}{Fernandez-Nieves, A.}, \bibinfo{year}{2023}.
\newblock \bibinfo{title}{Ant waves -- spontaneous activity waves in fire-ant
  columns}.
\newblock \bibinfo{journal}{Science Advances} \bibinfo{volume}{9},
  \bibinfo{pages}{eadd0635}.
\bibitem[{Asadchy et~al.(2020)Asadchy, Mirmoosa, Diaz-Rubio, Fan and
  Tretyakov}]{asadchy2020}
\bibinfo{author}{Asadchy, V.S.}, \bibinfo{author}{Mirmoosa, M.S.},
  \bibinfo{author}{Diaz-Rubio, A.}, \bibinfo{author}{Fan, S.},
  \bibinfo{author}{Tretyakov, S.A.}, \bibinfo{year}{2020}.
\newblock \bibinfo{title}{Tutorial on electromagnetic nonreciprocity and its
  origins}.
\newblock \bibinfo{journal}{Proceedings of the IEEE} \bibinfo{volume}{108},
  \bibinfo{pages}{1684--1727}.
\bibitem[{Aubin et~al.(2024)Aubin, Buskohl, Vaia and Shepherd}]{aubin2024}
\bibinfo{author}{Aubin, C.A.}, \bibinfo{author}{Buskohl, P.R.},
  \bibinfo{author}{Vaia, R.A.}, \bibinfo{author}{Shepherd, R.F.},
  \bibinfo{year}{2024}.
\newblock \bibinfo{title}{Autonomous material systems}.
\newblock \bibinfo{journal}{MRS Bulletin} \bibinfo{volume}{49},
  \bibinfo{pages}{1070--1078}.
\bibitem[{Bao and Cui(2020)}]{bao2020}
\bibinfo{author}{Bao, L.}, \bibinfo{author}{Cui, T.J.}, \bibinfo{year}{2020}.
\newblock \bibinfo{title}{Tunable, reconfigurable, and programmable
  metamaterials}.
\newblock \bibinfo{journal}{Microwave and Optical Technology Letters}
  \bibinfo{volume}{62}, \bibinfo{pages}{9--32}.
\bibitem[{Baulin et~al.(2026)Baulin, F{\"u}chslin, Giacometti, Hauser and
  Werner}]{baulin2026}
\bibinfo{author}{Baulin, V.A.}, \bibinfo{author}{F{\"u}chslin, R.M.},
  \bibinfo{author}{Giacometti, A.}, \bibinfo{author}{Hauser, H.},
  \bibinfo{author}{Werner, M.}, \bibinfo{year}{2026}.
\newblock \bibinfo{title}{Material-based intelligence: {Autonomous} adaptation
  and embodied computation in physical substrates}.
\newblock \bibinfo{journal}{Advanced Intelligent Systems} \bibinfo{volume}{8},
  \bibinfo{pages}{e202501450}.
\bibitem[{Brandenbourger et~al.(2019)Brandenbourger, Locsin, Lerner and
  Coulais}]{brandenbourger2019}
\bibinfo{author}{Brandenbourger, M.}, \bibinfo{author}{Locsin, X.},
  \bibinfo{author}{Lerner, E.}, \bibinfo{author}{Coulais, C.},
  \bibinfo{year}{2019}.
\newblock \bibinfo{title}{Non-reciprocal robotic metamaterials}.
\newblock \bibinfo{journal}{Nature Communications} \bibinfo{volume}{10},
  \bibinfo{pages}{4608}.
\bibitem[{Casimir(1945)}]{casimir1945}
\bibinfo{author}{Casimir, H.B.G.}, \bibinfo{year}{1945}.
\newblock \bibinfo{title}{On {Onsager's} principle of microscopic
  reversibility}.
\newblock \bibinfo{journal}{Reviews of Modern Physics} \bibinfo{volume}{17},
  \bibinfo{pages}{343}.
\bibitem[{Chen et~al.(2026)Chen, Wu, Schott and Jovanova}]{chen2026}
\bibinfo{author}{Chen, Q.}, \bibinfo{author}{Wu, R.}, \bibinfo{author}{Schott,
  D.}, \bibinfo{author}{Jovanova, J.}, \bibinfo{year}{2026}.
\newblock \bibinfo{title}{Programmable structure with shape memory materials
  for soft robotics}.
\newblock \bibinfo{journal}{Smart Materials and Structures} .
\bibitem[{Chen et~al.(2021)Chen, Li, Scheibner, Vitelli and Huang}]{chen2021}
\bibinfo{author}{Chen, Y.}, \bibinfo{author}{Li, X.},
  \bibinfo{author}{Scheibner, C.}, \bibinfo{author}{Vitelli, V.},
  \bibinfo{author}{Huang, G.}, \bibinfo{year}{2021}.
\newblock \bibinfo{title}{Realization of active metamaterials with odd
  micropolar elasticity}.
\newblock \bibinfo{journal}{Nature Communications} \bibinfo{volume}{12},
  \bibinfo{pages}{5935}.
\bibitem[{Chopra et~al.(2017)Chopra, Notarstefano, Rice and
  Egerstedt}]{chopra2017}
\bibinfo{author}{Chopra, S.}, \bibinfo{author}{Notarstefano, G.},
  \bibinfo{author}{Rice, M.}, \bibinfo{author}{Egerstedt, M.},
  \bibinfo{year}{2017}.
\newblock \bibinfo{title}{A distributed version of the {Hungarian} method for
  multirobot assignment}.
\newblock \bibinfo{journal}{IEEE Transactions on Robotics}
  \bibinfo{volume}{33}, \bibinfo{pages}{932--947}.
\bibitem[{Correll et~al.(2014)Correll, {\"O}nal, Liang, Schoenfeld and
  Rus}]{correll2014}
\bibinfo{author}{Correll, N.}, \bibinfo{author}{{\"O}nal, {\c{C}}.D.},
  \bibinfo{author}{Liang, H.}, \bibinfo{author}{Schoenfeld, E.},
  \bibinfo{author}{Rus, D.}, \bibinfo{year}{2014}.
\newblock \bibinfo{title}{Soft autonomous materials---using active elasticity
  and embedded distributed computation}, in: \bibinfo{booktitle}{Experimental
  Robotics: The 12th International Symposium on Experimental Robotics},
  \bibinfo{publisher}{Springer}, \bibinfo{address}{Berlin, Heidelberg}. pp.
  \bibinfo{pages}{227--240}.
\bibitem[{Coulais et~al.(2021)Coulais, Fleury and {van Wezel}}]{coulais2021}
\bibinfo{author}{Coulais, C.}, \bibinfo{author}{Fleury, R.},
  \bibinfo{author}{{van Wezel}, J.}, \bibinfo{year}{2021}.
\newblock \bibinfo{title}{Topology and broken {Hermiticity}}.
\newblock \bibinfo{journal}{Nature Physics} \bibinfo{volume}{17},
  \bibinfo{pages}{9--13}.
\bibitem[{Coulais et~al.(2017)Coulais, Sounas and Alu}]{coulais2017}
\bibinfo{author}{Coulais, C.}, \bibinfo{author}{Sounas, D.},
  \bibinfo{author}{Alu, A.}, \bibinfo{year}{2017}.
\newblock \bibinfo{title}{Static non-reciprocity in mechanical metamaterials}.
\newblock \bibinfo{journal}{Nature} \bibinfo{volume}{542},
  \bibinfo{pages}{461--464}.
\bibitem[{Cui et~al.(2014)Cui, Qi, Wan, Zhao and Cheng}]{cui2014}
\bibinfo{author}{Cui, T.J.}, \bibinfo{author}{Qi, M.Q.}, \bibinfo{author}{Wan,
  X.}, \bibinfo{author}{Zhao, J.}, \bibinfo{author}{Cheng, Q.},
  \bibinfo{year}{2014}.
\newblock \bibinfo{title}{Coding metamaterials, digital metamaterials and
  programmable metamaterials}.
\newblock \bibinfo{journal}{Light: Science \& Applications}
  \bibinfo{volume}{3}, \bibinfo{pages}{e218}.
\bibitem[{Deneubourg and Goss(1989)}]{deneubourg1989}
\bibinfo{author}{Deneubourg, J.L.}, \bibinfo{author}{Goss, S.},
  \bibinfo{year}{1989}.
\newblock \bibinfo{title}{Collective patterns and decision-making}.
\newblock \bibinfo{journal}{Ethology Ecology \& Evolution} \bibinfo{volume}{1},
  \bibinfo{pages}{295--311}.
\bibitem[{Dey et~al.(2010)Dey, Hirani and Krishnamoorthy}]{dey2010}
\bibinfo{author}{Dey, T.K.}, \bibinfo{author}{Hirani, A.N.},
  \bibinfo{author}{Krishnamoorthy, B.}, \bibinfo{year}{2010}.
\newblock \bibinfo{title}{Optimal homologous cycles, total unimodularity, and
  linear programming}, in: \bibinfo{booktitle}{Proceedings of the Forty-Second
  ACM Symposium on Theory of Computing}, pp. \bibinfo{pages}{221--230}.
\bibitem[{Dong et~al.(2020)Dong, Ha, Jung and Kim}]{dong2020}
\bibinfo{author}{Dong, J.G.}, \bibinfo{author}{Ha, S.Y.},
  \bibinfo{author}{Jung, J.}, \bibinfo{author}{Kim, D.}, \bibinfo{year}{2020}.
\newblock \bibinfo{title}{On the stochastic flocking of the {Cucker-Smale}
  flock with randomly switching topologies}.
\newblock \bibinfo{journal}{SIAM Journal on Control and Optimization}
  \bibinfo{volume}{58}, \bibinfo{pages}{2332--2353}.
\bibitem[{{El Helou} et~al.(2024){El Helou}, Hyatt, Buskohl and
  Harne}]{elhelou2024}
\bibinfo{author}{{El Helou}, C.}, \bibinfo{author}{Hyatt, L.P.},
  \bibinfo{author}{Buskohl, P.R.}, \bibinfo{author}{Harne, R.L.},
  \bibinfo{year}{2024}.
\newblock \bibinfo{title}{Intelligent electroactive material systems with
  self-adaptive mechanical memory and sequential logic}.
\newblock \bibinfo{journal}{Proceedings of the National Academy of Sciences}
  \bibinfo{volume}{121}, \bibinfo{pages}{e2317340121}.
\bibitem[{Fang et~al.(2022)Fang, Wen, Cheng, Yu, Zhang and Gumbsch}]{fang2022}
\bibinfo{author}{Fang, X.}, \bibinfo{author}{Wen, J.}, \bibinfo{author}{Cheng,
  L.}, \bibinfo{author}{Yu, D.}, \bibinfo{author}{Zhang, H.},
  \bibinfo{author}{Gumbsch, P.}, \bibinfo{year}{2022}.
\newblock \bibinfo{title}{Programmable gear-based mechanical metamaterials}.
\newblock \bibinfo{journal}{Nature Materials} \bibinfo{volume}{21},
  \bibinfo{pages}{869--876}.
\bibitem[{Fiedler(1973)}]{fiedler1973}
\bibinfo{author}{Fiedler, M.}, \bibinfo{year}{1973}.
\newblock \bibinfo{title}{Algebraic connectivity of graphs}.
\newblock \bibinfo{journal}{Czechoslovak Mathematical Journal}
  \bibinfo{volume}{23}, \bibinfo{pages}{298--305}.
\bibitem[{Florijn et~al.(2014)Florijn, Coulais and {Van Hecke}}]{florijn2014}
\bibinfo{author}{Florijn, B.}, \bibinfo{author}{Coulais, C.},
  \bibinfo{author}{{Van Hecke}, M.}, \bibinfo{year}{2014}.
\newblock \bibinfo{title}{Programmable mechanical metamaterials}.
\newblock \bibinfo{journal}{Physical Review Letters} \bibinfo{volume}{113},
  \bibinfo{pages}{175503}.
\bibitem[{Fruchart et~al.(2021)Fruchart, Hanai, Littlewood and
  Vitelli}]{fruchart2021}
\bibinfo{author}{Fruchart, M.}, \bibinfo{author}{Hanai, R.},
  \bibinfo{author}{Littlewood, P.B.}, \bibinfo{author}{Vitelli, V.},
  \bibinfo{year}{2021}.
\newblock \bibinfo{title}{Non-reciprocal phase transitions}.
\newblock \bibinfo{journal}{Nature} \bibinfo{volume}{592},
  \bibinfo{pages}{363--369}.
\bibitem[{Gibson and Ashby(1997)}]{gibson1997}
\bibinfo{author}{Gibson, L.J.}, \bibinfo{author}{Ashby, M.F.},
  \bibinfo{year}{1997}.
\newblock \bibinfo{title}{Cellular Solids: Structure and Properties}.
\newblock \bibinfo{edition}{2nd} ed., \bibinfo{publisher}{Cambridge University
  Press}, \bibinfo{address}{Cambridge}.
\bibitem[{Godsil and Royle(2001)}]{godsil2001}
\bibinfo{author}{Godsil, C.}, \bibinfo{author}{Royle, G.},
  \bibinfo{year}{2001}.
\newblock \bibinfo{title}{Algebraic Graph Theory}.
\newblock \bibinfo{publisher}{Springer}, \bibinfo{address}{New York}.
\bibitem[{Goldman and {Zeb Rocklin}(2024)}]{goldman2024}
\bibinfo{author}{Goldman, D.I.}, \bibinfo{author}{{Zeb Rocklin}, D.},
  \bibinfo{year}{2024}.
\newblock \bibinfo{title}{Robot swarms meet soft matter physics}.
\newblock \bibinfo{journal}{Science Robotics} \bibinfo{volume}{9},
  \bibinfo{pages}{eadn6035}.
\bibitem[{Gregg et~al.(2024)Gregg, Catanoso, Formoso, Kostitsyna, Ochalek,
  Olatunde et~al.}]{gregg2024}
\bibinfo{author}{Gregg, C.E.}, \bibinfo{author}{Catanoso, D.},
  \bibinfo{author}{Formoso, O.I.B.}, \bibinfo{author}{Kostitsyna, I.},
  \bibinfo{author}{Ochalek, M.E.}, \bibinfo{author}{Olatunde, T.J.}, et~al.,
  \bibinfo{year}{2024}.
\newblock \bibinfo{title}{Ultralight, strong, and self-reprogrammable
  mechanical metamaterials}.
\newblock \bibinfo{journal}{Science Robotics} \bibinfo{volume}{9},
  \bibinfo{pages}{eadi2746}.
\bibitem[{Hagan and Baskaran(2016)}]{hagan2016}
\bibinfo{author}{Hagan, M.F.}, \bibinfo{author}{Baskaran, A.},
  \bibinfo{year}{2016}.
\newblock \bibinfo{title}{Emergent self-organization in active materials}.
\newblock \bibinfo{journal}{Current Opinion in Cell Biology}
  \bibinfo{volume}{38}, \bibinfo{pages}{74--80}.
\bibitem[{Haghpanah et~al.(2016)Haghpanah, Ebrahimi, Mousanezhad, Hopkins and
  Vaziri}]{haghpanah2016}
\bibinfo{author}{Haghpanah, B.}, \bibinfo{author}{Ebrahimi, H.},
  \bibinfo{author}{Mousanezhad, D.}, \bibinfo{author}{Hopkins, J.},
  \bibinfo{author}{Vaziri, A.}, \bibinfo{year}{2016}.
\newblock \bibinfo{title}{Programmable elastic metamaterials}.
\newblock \bibinfo{journal}{Advanced Engineering Materials}
  \bibinfo{volume}{18}, \bibinfo{pages}{643--649}.
\bibitem[{Hill(1963)}]{hill1963}
\bibinfo{author}{Hill, R.}, \bibinfo{year}{1963}.
\newblock \bibinfo{title}{Elastic properties of reinforced solids: {Some}
  theoretical principles}.
\newblock \bibinfo{journal}{Journal of the Mechanics and Physics of Solids}
  \bibinfo{volume}{11}, \bibinfo{pages}{357--372}.
\bibitem[{Huang et~al.(2025)Huang, Qi and Srigrarom}]{huang2025}
\bibinfo{author}{Huang, D.}, \bibinfo{author}{Qi, R.},
  \bibinfo{author}{Srigrarom, S.}, \bibinfo{year}{2025}.
\newblock \bibinfo{title}{Hungarian algorithm for drone swarm formation and
  energy optimization}, in: \bibinfo{booktitle}{2025 17th International
  Conference on Information Technology and Electrical Engineering (ICITEE)},
  \bibinfo{organization}{IEEE}. pp. \bibinfo{pages}{1--6}.
\bibitem[{Huang et~al.(2018)Huang, Ding, Zhou, Jin and Hao}]{huang2018}
\bibinfo{author}{Huang, L.}, \bibinfo{author}{Ding, Y.}, \bibinfo{author}{Zhou,
  M.}, \bibinfo{author}{Jin, Y.}, \bibinfo{author}{Hao, K.},
  \bibinfo{year}{2018}.
\newblock \bibinfo{title}{Multiple-solution optimization strategy for
  multirobot task allocation}.
\newblock \bibinfo{journal}{IEEE Transactions on Systems, Man, and Cybernetics:
  Systems} \bibinfo{volume}{50}, \bibinfo{pages}{4283--4294}.
\bibitem[{Ismail and Sun(2017)}]{ismail2017}
\bibinfo{author}{Ismail, S.}, \bibinfo{author}{Sun, L.}, \bibinfo{year}{2017}.
\newblock \bibinfo{title}{Decentralized {Hungarian-based} approach for fast and
  scalable task allocation}, in: \bibinfo{booktitle}{2017 International
  Conference on Unmanned Aircraft Systems (ICUAS)},
  \bibinfo{organization}{IEEE}. pp. \bibinfo{pages}{23--28}.
\bibitem[{Jonker and Volgenant(1986)}]{jonker1986}
\bibinfo{author}{Jonker, R.}, \bibinfo{author}{Volgenant, T.},
  \bibinfo{year}{1986}.
\newblock \bibinfo{title}{Improving the {Hungarian} assignment algorithm}.
\newblock \bibinfo{journal}{Operations Research Letters} \bibinfo{volume}{5},
  \bibinfo{pages}{171--175}.
\bibitem[{Julius et~al.(2008)Julius, Hal{\'a}sz, Sakar, Rubin, Kumar and
  Pappas}]{julius2008}
\bibinfo{author}{Julius, A.A.}, \bibinfo{author}{Hal{\'a}sz, {\'A}.},
  \bibinfo{author}{Sakar, M.S.}, \bibinfo{author}{Rubin, H.},
  \bibinfo{author}{Kumar, V.}, \bibinfo{author}{Pappas, G.J.},
  \bibinfo{year}{2008}.
\newblock \bibinfo{title}{Stochastic modeling and control of biological
  systems: the lactose regulation system of {Escherichia} coli}.
\newblock \bibinfo{journal}{IEEE Transactions on Automatic Control}
  \bibinfo{volume}{53}, \bibinfo{pages}{51--65}.
\bibitem[{Kuhn(1955)}]{kuhn1955}
\bibinfo{author}{Kuhn, H.W.}, \bibinfo{year}{1955}.
\newblock \bibinfo{title}{The {Hungarian} method for the assignment problem}.
\newblock \bibinfo{journal}{Naval Research Logistics Quarterly}
  \bibinfo{volume}{2}, \bibinfo{pages}{83--97}.
\bibitem[{Kumar et~al.(2008)Kumar, Milutinovic and Garg}]{kumar2008}
\bibinfo{author}{Kumar, M.}, \bibinfo{author}{Milutinovic, D.},
  \bibinfo{author}{Garg, D.P.}, \bibinfo{year}{2008}.
\newblock \bibinfo{title}{Role of stochasticity in self-organization of robotic
  swarms}, in: \bibinfo{booktitle}{2008 American Control Conference},
  \bibinfo{organization}{IEEE}. pp. \bibinfo{pages}{123--128}.
\bibitem[{Kume and Walker(2021)}]{kume2021}
\bibinfo{author}{Kume, A.}, \bibinfo{author}{Walker, S.G.},
  \bibinfo{year}{2021}.
\newblock \bibinfo{title}{The utility of clusters and a {Hungarian} clustering
  algorithm}.
\newblock \bibinfo{journal}{PLoS ONE} \bibinfo{volume}{16},
  \bibinfo{pages}{e0255174}.
\bibitem[{Liu et~al.(2025)Liu, Zhu, Gosrich, Yim and Raney}]{liu2025}
\bibinfo{author}{Liu, J.}, \bibinfo{author}{Zhu, X.}, \bibinfo{author}{Gosrich,
  W.}, \bibinfo{author}{Yim, M.}, \bibinfo{author}{Raney, J.R.},
  \bibinfo{year}{2025}.
\newblock \bibinfo{title}{Design of nondeterministic architected structures via
  bioinspired distributed agents}.
\newblock \bibinfo{journal}{Science Advances} \bibinfo{volume}{11},
  \bibinfo{pages}{eadu8260}.
\bibitem[{Losick and Desplan(2008)}]{losick2008}
\bibinfo{author}{Losick, R.}, \bibinfo{author}{Desplan, C.},
  \bibinfo{year}{2008}.
\newblock \bibinfo{title}{Stochasticity and cell fate}.
\newblock \bibinfo{journal}{Science} \bibinfo{volume}{320},
  \bibinfo{pages}{65--68}.
\bibitem[{Lukeman et~al.(2010)Lukeman, Li and Edelstein-Keshet}]{lukeman2010}
\bibinfo{author}{Lukeman, R.}, \bibinfo{author}{Li, Y.X.},
  \bibinfo{author}{Edelstein-Keshet, L.}, \bibinfo{year}{2010}.
\newblock \bibinfo{title}{Inferring individual rules from collective behavior}.
\newblock \bibinfo{journal}{Proceedings of the National Academy of Sciences}
  \bibinfo{volume}{107}, \bibinfo{pages}{12576--12580}.
\bibitem[{Markvicka et~al.(2018)Markvicka, Bartlett, Huang and
  Majidi}]{markvicka2018}
\bibinfo{author}{Markvicka, E.J.}, \bibinfo{author}{Bartlett, M.D.},
  \bibinfo{author}{Huang, X.}, \bibinfo{author}{Majidi, C.},
  \bibinfo{year}{2018}.
\newblock \bibinfo{title}{An autonomously electrically self-healing liquid
  metal--elastomer composite for robust soft-matter robotics and electronics}.
\newblock \bibinfo{journal}{Nature Materials} \bibinfo{volume}{17},
  \bibinfo{pages}{618--624}.
\bibitem[{Meng et~al.(2024)Meng, Yan and Wang}]{meng2024}
\bibinfo{author}{Meng, Z.}, \bibinfo{author}{Yan, H.}, \bibinfo{author}{Wang,
  Y.}, \bibinfo{year}{2024}.
\newblock \bibinfo{title}{Granular metamaterials with dynamic bond
  reconfiguration}.
\newblock \bibinfo{journal}{Science Advances} \bibinfo{volume}{10},
  \bibinfo{pages}{eadq7933}.
\bibitem[{Merindol and Walther(2017)}]{merindol2017}
\bibinfo{author}{Merindol, R.}, \bibinfo{author}{Walther, A.},
  \bibinfo{year}{2017}.
\newblock \bibinfo{title}{Materials learning from life: concepts for active,
  adaptive and autonomous molecular systems}.
\newblock \bibinfo{journal}{Chemical Society Reviews} \bibinfo{volume}{46},
  \bibinfo{pages}{5588--5619}.
\bibitem[{Mesbahi and Egerstedt(2010)}]{mesbahi2010}
\bibinfo{author}{Mesbahi, M.}, \bibinfo{author}{Egerstedt, M.},
  \bibinfo{year}{2010}.
\newblock \bibinfo{title}{Graph Theoretic Methods in Multiagent Networks}.
  volume~\bibinfo{volume}{33} of \textit{\bibinfo{series}{Princeton Series in
  Applied Mathematics}}.
\newblock \bibinfo{publisher}{Princeton University Press},
  \bibinfo{address}{Princeton}.
\bibitem[{Mills-Tettey et~al.(2007)Mills-Tettey, Stentz and
  Dias}]{millstettey2007}
\bibinfo{author}{Mills-Tettey, G.A.}, \bibinfo{author}{Stentz, A.},
  \bibinfo{author}{Dias, M.B.}, \bibinfo{year}{2007}.
\newblock \bibinfo{title}{The dynamic {Hungarian} algorithm for the assignment
  problem with changing costs}.
\newblock \bibinfo{type}{Technical Report} \bibinfo{number}{CMU-RI-TR-07-27}.
  Robotics Institute. \bibinfo{address}{Pittsburgh, PA}.
\bibitem[{Mindlin and Tiersten(1962)}]{mindlin1962}
\bibinfo{author}{Mindlin, R.D.}, \bibinfo{author}{Tiersten, H.},
  \bibinfo{year}{1962}.
\newblock \bibinfo{title}{Effects of couple-stresses in linear elasticity}.
\newblock \bibinfo{journal}{Archive for Rational Mechanics and Analysis}
  \bibinfo{volume}{11}, \bibinfo{pages}{415--448}.
\bibitem[{Miskin(2026)}]{miskin2026}
\bibinfo{author}{Miskin, M.Z.}, \bibinfo{year}{2026}.
\newblock \bibinfo{title}{Robotic matter}.
\newblock \bibinfo{journal}{Annual Review of Condensed Matter Physics}
  \bibinfo{volume}{17}, \bibinfo{pages}{449--471}.
\bibitem[{Montazeri et~al.(2025)Montazeri, Rahimi and Park}]{montazeri2025}
\bibinfo{author}{Montazeri, A.}, \bibinfo{author}{Rahimi, M.},
  \bibinfo{author}{Park, H.S.}, \bibinfo{year}{2025}.
\newblock \bibinfo{title}{Non-reciprocity and asymmetric elasticity in twisting
  chiral metamaterials}.
\newblock \bibinfo{journal}{International Journal of Mechanical Sciences}
  \bibinfo{volume}{287}, \bibinfo{pages}{109990}.
\bibitem[{Moreno and Mossio(2015)}]{moreno2015}
\bibinfo{author}{Moreno, A.}, \bibinfo{author}{Mossio, M.},
  \bibinfo{year}{2015}.
\newblock \bibinfo{title}{Biological Autonomy: A Philosophical and Theoretical
  Enquiry}.
\newblock \bibinfo{publisher}{Springer}, \bibinfo{address}{New York}.
\bibitem[{Nash et~al.(2015)Nash, Kleckner, Read, Vitelli, Turner and
  Irvine}]{nash2015}
\bibinfo{author}{Nash, L.M.}, \bibinfo{author}{Kleckner, D.},
  \bibinfo{author}{Read, A.}, \bibinfo{author}{Vitelli, V.},
  \bibinfo{author}{Turner, A.M.}, \bibinfo{author}{Irvine, W.T.},
  \bibinfo{year}{2015}.
\newblock \bibinfo{title}{Topological mechanics of gyroscopic metamaterials}.
\newblock \bibinfo{journal}{Proceedings of the National Academy of Sciences}
  \bibinfo{volume}{112}, \bibinfo{pages}{14495--14500}.
\bibitem[{Nassar et~al.(2017)Nassar, Chen, Norris, Haberman and
  Huang}]{nassar2017}
\bibinfo{author}{Nassar, H.}, \bibinfo{author}{Chen, H.},
  \bibinfo{author}{Norris, A.N.}, \bibinfo{author}{Haberman, M.R.},
  \bibinfo{author}{Huang, G.L.}, \bibinfo{year}{2017}.
\newblock \bibinfo{title}{Non-reciprocal wave propagation in modulated elastic
  metamaterials}.
\newblock \bibinfo{journal}{Proceedings of the Royal Society A: Mathematical,
  Physical and Engineering Sciences} \bibinfo{volume}{473},
  \bibinfo{pages}{20170188}.
\bibitem[{Nassar et~al.(2020)Nassar, Yousefzadeh, Fleury, Ruzzene, Al{\`u},
  Daraio, Norris, Huang and Haberman}]{nassar2020}
\bibinfo{author}{Nassar, H.}, \bibinfo{author}{Yousefzadeh, B.},
  \bibinfo{author}{Fleury, R.}, \bibinfo{author}{Ruzzene, M.},
  \bibinfo{author}{Al{\`u}, A.}, \bibinfo{author}{Daraio, C.},
  \bibinfo{author}{Norris, A.N.}, \bibinfo{author}{Huang, G.},
  \bibinfo{author}{Haberman, M.R.}, \bibinfo{year}{2020}.
\newblock \bibinfo{title}{Nonreciprocity in acoustic and elastic materials}.
\newblock \bibinfo{journal}{Nature Reviews Materials} \bibinfo{volume}{5},
  \bibinfo{pages}{667--685}.
\bibitem[{Nguyen et~al.(2026)Nguyen, Shvydkoy and Tan}]{nguyen2026}
\bibinfo{author}{Nguyen, V.}, \bibinfo{author}{Shvydkoy, R.},
  \bibinfo{author}{Tan, C.}, \bibinfo{year}{2026}.
\newblock \bibinfo{title}{Mean-field limits of deterministic and stochastic
  flocking models with nonlinear velocity alignment}.
\newblock \bibinfo{journal}{Advances in Mathematics} \bibinfo{volume}{494},
  \bibinfo{pages}{110929}.
\bibitem[{{{\'O} Laighl{\'e}is} et~al.(2018){{\'O} Laighl{\'e}is}, Evans and
  Blythe}]{olaighleis2018}
\bibinfo{author}{{{\'O} Laighl{\'e}is}, E.}, \bibinfo{author}{Evans, M.R.},
  \bibinfo{author}{Blythe, R.A.}, \bibinfo{year}{2018}.
\newblock \bibinfo{title}{Minimal stochastic field equations for
  one-dimensional flocking}.
\newblock \bibinfo{journal}{Physical Review E} \bibinfo{volume}{98},
  \bibinfo{pages}{062127}.
\bibitem[{Okubo(1986)}]{okubo1986}
\bibinfo{author}{Okubo, A.}, \bibinfo{year}{1986}.
\newblock \bibinfo{title}{Dynamical aspects of animal grouping: swarms,
  schools, flocks, and herds}.
\newblock \bibinfo{journal}{Advances in Biophysics} \bibinfo{volume}{22},
  \bibinfo{pages}{1--94}.
\bibitem[{Olfati-Saber(2006)}]{olfatisaber2006}
\bibinfo{author}{Olfati-Saber, R.}, \bibinfo{year}{2006}.
\newblock \bibinfo{title}{Flocking for multi-agent dynamic systems:
  {Algorithms} and theory}.
\newblock \bibinfo{journal}{IEEE Transactions on Automatic Control}
  \bibinfo{volume}{51}, \bibinfo{pages}{401--420}.
\bibitem[{Olfati-Saber et~al.(2007)Olfati-Saber, Fax and
  Murray}]{olfatisaber2007}
\bibinfo{author}{Olfati-Saber, R.}, \bibinfo{author}{Fax, J.A.},
  \bibinfo{author}{Murray, R.M.}, \bibinfo{year}{2007}.
\newblock \bibinfo{title}{Consensus and cooperation in networked multi-agent
  systems}.
\newblock \bibinfo{journal}{Proceedings of the IEEE} \bibinfo{volume}{95},
  \bibinfo{pages}{215--233}.
\bibitem[{Olfati-Saber and Murray(2003)}]{olfatisaber2003}
\bibinfo{author}{Olfati-Saber, R.}, \bibinfo{author}{Murray, R.M.},
  \bibinfo{year}{2003}.
\newblock \bibinfo{title}{Flocking with obstacle avoidance: {Cooperation} with
  limited communication in mobile networks}, in: \bibinfo{booktitle}{42nd IEEE
  International Conference on Decision and Control (IEEE Cat. No. 03CH37475)},
  \bibinfo{organization}{IEEE}. pp. \bibinfo{pages}{2022--2028}.
\bibitem[{Olfati-Saber and Murray(2004)}]{olfatisaber2004}
\bibinfo{author}{Olfati-Saber, R.}, \bibinfo{author}{Murray, R.M.},
  \bibinfo{year}{2004}.
\newblock \bibinfo{title}{Consensus problems in networks of agents with
  switching topology and time-delays}.
\newblock \bibinfo{journal}{IEEE Transactions on Automatic Control}
  \bibinfo{volume}{49}, \bibinfo{pages}{1520--1533}.
\bibitem[{Olson(1997)}]{olson1997}
\bibinfo{author}{Olson, G.B.}, \bibinfo{year}{1997}.
\newblock \bibinfo{title}{Computational design of hierarchically structured
  materials}.
\newblock \bibinfo{journal}{Science} \bibinfo{volume}{277},
  \bibinfo{pages}{1237--1242}.
\bibitem[{Orrego et~al.(2020)Orrego, Chen, Krekora, Hou, Jeon, Pittman
  et~al.}]{orrego2020}
\bibinfo{author}{Orrego, S.}, \bibinfo{author}{Chen, Z.},
  \bibinfo{author}{Krekora, U.}, \bibinfo{author}{Hou, D.},
  \bibinfo{author}{Jeon, S.Y.}, \bibinfo{author}{Pittman, M.}, et~al.,
  \bibinfo{year}{2020}.
\newblock \bibinfo{title}{Bioinspired materials with self-adaptable mechanical
  properties}.
\newblock \bibinfo{journal}{Advanced Materials} \bibinfo{volume}{32},
  \bibinfo{pages}{1906970}.
\bibitem[{Otsuka and Wayman(1999)}]{otsuka1999}
\bibinfo{editor}{Otsuka, K.}, \bibinfo{editor}{Wayman, C.M.} (Eds.),
  \bibinfo{year}{1999}.
\newblock \bibinfo{title}{Shape Memory Materials}.
\newblock \bibinfo{publisher}{Cambridge University Press}.
\bibitem[{Palacci et~al.(2013)Palacci, Sacanna, Steinberg, Pine and
  Chaikin}]{palacci2013}
\bibinfo{author}{Palacci, J.}, \bibinfo{author}{Sacanna, S.},
  \bibinfo{author}{Steinberg, A.P.}, \bibinfo{author}{Pine, D.J.},
  \bibinfo{author}{Chaikin, P.M.}, \bibinfo{year}{2013}.
\newblock \bibinfo{title}{Living crystals of light-activated colloidal
  surfers}.
\newblock \bibinfo{journal}{Science} \bibinfo{volume}{339},
  \bibinfo{pages}{936--940}.
\bibitem[{Parrish et~al.(2002)Parrish, Viscido and Grunbaum}]{parrish2002}
\bibinfo{author}{Parrish, J.K.}, \bibinfo{author}{Viscido, S.V.},
  \bibinfo{author}{Grunbaum, D.}, \bibinfo{year}{2002}.
\newblock \bibinfo{title}{Self-organized fish schools: an examination of
  emergent properties}.
\newblock \bibinfo{journal}{The Biological Bulletin} \bibinfo{volume}{202},
  \bibinfo{pages}{296--305}.
\bibitem[{Partridge(1982)}]{partridge1982}
\bibinfo{author}{Partridge, B.L.}, \bibinfo{year}{1982}.
\newblock \bibinfo{title}{The structure and function of fish schools}.
\newblock \bibinfo{journal}{Scientific American} \bibinfo{volume}{246},
  \bibinfo{pages}{114--123}.
\bibitem[{Penrose(2003)}]{penrose2003}
\bibinfo{author}{Penrose, M.}, \bibinfo{year}{2003}.
\newblock \bibinfo{title}{Random Geometric Graphs}.
\newblock \bibinfo{publisher}{Oxford University Press},
  \bibinfo{address}{Oxford}.
\bibitem[{Pfeifer et~al.(2007)Pfeifer, Lungarella and Iida}]{pfeifer2007}
\bibinfo{author}{Pfeifer, R.}, \bibinfo{author}{Lungarella, M.},
  \bibinfo{author}{Iida, F.}, \bibinfo{year}{2007}.
\newblock \bibinfo{title}{Self-organization, embodiment, and biologically
  inspired robotics}.
\newblock \bibinfo{journal}{Science} \bibinfo{volume}{318},
  \bibinfo{pages}{1088--1093}.
\bibitem[{Reynolds(2023)}]{reynolds2023}
\bibinfo{author}{Reynolds, A.M.}, \bibinfo{year}{2023}.
\newblock \bibinfo{title}{Stochasticity may generate coherent motion in bird
  flocks}.
\newblock \bibinfo{journal}{Physical Biology} \bibinfo{volume}{20},
  \bibinfo{pages}{025002}.
\bibitem[{Reynolds et~al.(2022)Reynolds, McIvor, Thornton, Yang and
  Ouellette}]{reynolds2022}
\bibinfo{author}{Reynolds, A.M.}, \bibinfo{author}{McIvor, G.E.},
  \bibinfo{author}{Thornton, A.}, \bibinfo{author}{Yang, P.},
  \bibinfo{author}{Ouellette, N.T.}, \bibinfo{year}{2022}.
\newblock \bibinfo{title}{Stochastic modelling of bird flocks: accounting for
  the cohesiveness of collective motion}.
\newblock \bibinfo{journal}{Journal of the Royal Society Interface}
  \bibinfo{volume}{19}, \bibinfo{pages}{20210745}.
\bibitem[{Reynolds(1987)}]{reynolds1987}
\bibinfo{author}{Reynolds, C.W.}, \bibinfo{year}{1987}.
\newblock \bibinfo{title}{Flocks, herds and schools: {A} distributed behavioral
  model}, in: \bibinfo{booktitle}{Proceedings of the 14th Annual Conference on
  Computer Graphics and Interactive Techniques}, pp. \bibinfo{pages}{25--34}.
\bibitem[{Rubenstein et~al.(2014a)Rubenstein, Ahler, Hoff, Cabrera and
  Nagpal}]{rubenstein2014b}
\bibinfo{author}{Rubenstein, M.}, \bibinfo{author}{Ahler, C.},
  \bibinfo{author}{Hoff, N.}, \bibinfo{author}{Cabrera, A.},
  \bibinfo{author}{Nagpal, R.}, \bibinfo{year}{2014}a.
\newblock \bibinfo{title}{Kilobot: {A} low cost robot with scalable operations
  designed for collective behaviors}.
\newblock \bibinfo{journal}{Robotics and Autonomous Systems}
  \bibinfo{volume}{62}, \bibinfo{pages}{966--975}.
\bibitem[{Rubenstein et~al.(2014b)Rubenstein, Cornejo and
  Nagpal}]{rubenstein2014a}
\bibinfo{author}{Rubenstein, M.}, \bibinfo{author}{Cornejo, A.},
  \bibinfo{author}{Nagpal, R.}, \bibinfo{year}{2014}b.
\newblock \bibinfo{title}{Programmable self-assembly in a thousand-robot
  swarm}.
\newblock \bibinfo{journal}{Science} \bibinfo{volume}{345},
  \bibinfo{pages}{795--799}.
\bibitem[{Ruiz-Mirazo and Moreno(2004)}]{ruizmirazo2004}
\bibinfo{author}{Ruiz-Mirazo, K.}, \bibinfo{author}{Moreno, A.},
  \bibinfo{year}{2004}.
\newblock \bibinfo{title}{Basic autonomy as a fundamental step in the synthesis
  of life}.
\newblock \bibinfo{journal}{Artificial Life} \bibinfo{volume}{10},
  \bibinfo{pages}{235--259}.
\bibitem[{Samiei et~al.(2019)Samiei, Ismail and Sun}]{samiei2019}
\bibinfo{author}{Samiei, A.}, \bibinfo{author}{Ismail, S.},
  \bibinfo{author}{Sun, L.}, \bibinfo{year}{2019}.
\newblock \bibinfo{title}{Cluster-based {Hungarian} approach to task allocation
  for unmanned aerial vehicles}, in: \bibinfo{booktitle}{2019 IEEE National
  Aerospace and Electronics Conference (NAECON)}, \bibinfo{organization}{IEEE}.
  pp. \bibinfo{pages}{148--154}.
\bibitem[{Scheibner et~al.(2020)Scheibner, Souslov, Banerjee, Sur{\'o}wka,
  Irvine and Vitelli}]{scheibner2020b}
\bibinfo{author}{Scheibner, C.}, \bibinfo{author}{Souslov, A.},
  \bibinfo{author}{Banerjee, D.}, \bibinfo{author}{Sur{\'o}wka, P.},
  \bibinfo{author}{Irvine, W.T.}, \bibinfo{author}{Vitelli, V.},
  \bibinfo{year}{2020}.
\newblock \bibinfo{title}{Odd elasticity}.
\newblock \bibinfo{journal}{Nature Physics} \bibinfo{volume}{16},
  \bibinfo{pages}{475--480}.
\bibitem[{Shaat(2018)}]{shaat2018c}
\bibinfo{author}{Shaat, M.}, \bibinfo{year}{2018}.
\newblock \bibinfo{title}{Effects of processing conditions on microstructure
  and mechanical properties of equal-channel-angular-pressed titanium}.
\newblock \bibinfo{journal}{Materials Science and Technology}
  \bibinfo{volume}{34}, \bibinfo{pages}{1149--1167}.
\bibitem[{Shaat(2020)}]{shaat2020d}
\bibinfo{author}{Shaat, M.}, \bibinfo{year}{2020}.
\newblock \bibinfo{title}{Nonreciprocal elasticity and the realization of
  static and dynamic nonreciprocity}.
\newblock \bibinfo{journal}{Scientific Reports} \bibinfo{volume}{10},
  \bibinfo{pages}{21676}.
\bibitem[{Shaat et~al.(2020a)Shaat, Fathy and Wagih}]{shaat2020e}
\bibinfo{author}{Shaat, M.}, \bibinfo{author}{Fathy, A.},
  \bibinfo{author}{Wagih, A.}, \bibinfo{year}{2020}a.
\newblock \bibinfo{title}{Correlation between grain boundary evolution and
  mechanical properties of ultrafine-grained metals}.
\newblock \bibinfo{journal}{Mechanics of Materials} \bibinfo{volume}{143},
  \bibinfo{pages}{103321}.
\bibitem[{Shaat and Gao(2024)}]{shaat2024a}
\bibinfo{author}{Shaat, M.}, \bibinfo{author}{Gao, X.L.}, \bibinfo{year}{2024}.
\newblock \bibinfo{title}{Topological boundary states in micropolar gyroelastic
  continua}.
\newblock \bibinfo{journal}{Mechanics of Materials} \bibinfo{volume}{192},
  \bibinfo{pages}{104902}.
\bibitem[{Shaat and Gao(2025a)}]{shaat2025a}
\bibinfo{author}{Shaat, M.}, \bibinfo{author}{Gao, X.L.},
  \bibinfo{year}{2025}a.
\newblock \bibinfo{title}{Constitutive modeling of mechanical coupling in {2D}
  biomimetic metamaterials with chirality, anisotropy and non-centrosymmetry}.
\newblock \bibinfo{journal}{International Journal of Mechanical Sciences}
  \bibinfo{volume}{295}, \bibinfo{pages}{110276}.
\bibitem[{Shaat and Gao(2025b)}]{shaat2025b}
\bibinfo{author}{Shaat, M.}, \bibinfo{author}{Gao, X.L.},
  \bibinfo{year}{2025}b.
\newblock \bibinfo{title}{Topology optimization of {2D} chiral metamaterials
  with dilatation-shear and shear-shear coupling capabilities}.
\newblock \bibinfo{journal}{International Journal of Engineering Science}
  \bibinfo{volume}{217}, \bibinfo{pages}{104367}.
\bibitem[{Shaat and Gao(2026)}]{shaat2026b}
\bibinfo{author}{Shaat, M.}, \bibinfo{author}{Gao, X.L.}, \bibinfo{year}{2026}.
\newblock \bibinfo{title}{Hybrid multilayer perceptron-convolutional neural
  network model for non-iterative topology optimization of {2D} chiral
  metamaterials with shear-shear coupling}.
\newblock \bibinfo{journal}{International Journal of Mechanical Sciences} ,
  \bibinfo{pages}{111667}.
\bibitem[{Shaat et~al.(2020b)Shaat, Moubarez, Khan, Khan and
  Alzo'ubi}]{shaat2020a}
\bibinfo{author}{Shaat, M.}, \bibinfo{author}{Moubarez, M.A.},
  \bibinfo{author}{Khan, M.O.}, \bibinfo{author}{Khan, M.A.},
  \bibinfo{author}{Alzo'ubi, A.}, \bibinfo{year}{2020}b.
\newblock \bibinfo{title}{Metamaterials with giant and tailorable nonreciprocal
  elastic moduli}.
\newblock \bibinfo{journal}{Physical Review Applied} \bibinfo{volume}{14},
  \bibinfo{pages}{014005}.
\bibitem[{Shaat and Park(2023)}]{shaat2023}
\bibinfo{author}{Shaat, M.}, \bibinfo{author}{Park, H.S.},
  \bibinfo{year}{2023}.
\newblock \bibinfo{title}{Chiral nonreciprocal elasticity and mechanical
  activity}.
\newblock \bibinfo{journal}{Journal of the Mechanics and Physics of Solids}
  \bibinfo{volume}{171}, \bibinfo{pages}{105163}.
\bibitem[{Shaat and Park(2026)}]{shaat2026c}
\bibinfo{author}{Shaat, M.}, \bibinfo{author}{Park, H.S.},
  \bibinfo{year}{2026}.
\newblock \bibinfo{title}{Manipulation of {non-Hermitian} skin effect in
  elastic media}.
\newblock \bibinfo{journal}{International Journal of Mechanical Sciences} ,
  \bibinfo{pages}{111842}.
\bibitem[{Shankar et~al.(2022)Shankar, Souslov, Bowick, Marchetti and
  Vitelli}]{shankar2022}
\bibinfo{author}{Shankar, S.}, \bibinfo{author}{Souslov, A.},
  \bibinfo{author}{Bowick, M.J.}, \bibinfo{author}{Marchetti, M.C.},
  \bibinfo{author}{Vitelli, V.}, \bibinfo{year}{2022}.
\newblock \bibinfo{title}{Topological active matter}.
\newblock \bibinfo{journal}{Nature Reviews Physics} \bibinfo{volume}{4},
  \bibinfo{pages}{380--398}.
\bibitem[{Shimoyama et~al.(1996)Shimoyama, Sugawara, Mizuguchi, Hayakawa and
  Sano}]{shimoyama1996}
\bibinfo{author}{Shimoyama, N.}, \bibinfo{author}{Sugawara, K.},
  \bibinfo{author}{Mizuguchi, T.}, \bibinfo{author}{Hayakawa, Y.},
  \bibinfo{author}{Sano, M.}, \bibinfo{year}{1996}.
\newblock \bibinfo{title}{Collective motion in a system of motile elements}.
\newblock \bibinfo{journal}{Physical Review Letters} \bibinfo{volume}{76},
  \bibinfo{pages}{3870--3873}.
\bibitem[{Shobeiri and Xie(2025)}]{shobeiri2025}
\bibinfo{author}{Shobeiri, V.}, \bibinfo{author}{Xie, Y.M.},
  \bibinfo{year}{2025}.
\newblock \bibinfo{title}{Topology optimization for creating nonreciprocal
  compliant mechanisms: {Numerical} and experimental investigations}.
\newblock \bibinfo{journal}{Extreme Mechanics Letters} \bibinfo{volume}{77},
  \bibinfo{pages}{102345}.
\bibitem[{Sigmund(2001)}]{sigmund2001}
\bibinfo{author}{Sigmund, O.}, \bibinfo{year}{2001}.
\newblock \bibinfo{title}{A 99 line topology optimization code written in
  {Matlab}}.
\newblock \bibinfo{journal}{Structural and Multidisciplinary Optimization}
  \bibinfo{volume}{21}, \bibinfo{pages}{120--127}.
\bibitem[{Sigmund and Maute(2013)}]{sigmund2013}
\bibinfo{author}{Sigmund, O.}, \bibinfo{author}{Maute, K.},
  \bibinfo{year}{2013}.
\newblock \bibinfo{title}{Topology optimization approaches: {A} comparative
  review}.
\newblock \bibinfo{journal}{Structural and Multidisciplinary Optimization}
  \bibinfo{volume}{48}, \bibinfo{pages}{1031--1055}.
\bibitem[{Sui et~al.(2017)Sui, Pu and Yi}]{sui2017}
\bibinfo{author}{Sui, Z.}, \bibinfo{author}{Pu, Z.}, \bibinfo{author}{Yi, J.},
  \bibinfo{year}{2017}.
\newblock \bibinfo{title}{Optimal {UAVs} formation transformation strategy
  based on task assignment and {Particle} {Swarm} {Optimization}}, in:
  \bibinfo{booktitle}{2017 IEEE International Conference on Mechatronics and
  Automation (ICMA)}, \bibinfo{organization}{IEEE}. pp.
  \bibinfo{pages}{1804--1809}.
\bibitem[{Sun et~al.(2012)Sun, Huang, Ding, Zhao, Wang, Purnawali and
  Tang}]{sun2012}
\bibinfo{author}{Sun, L.}, \bibinfo{author}{Huang, W.M.},
  \bibinfo{author}{Ding, Z.}, \bibinfo{author}{Zhao, Y.},
  \bibinfo{author}{Wang, C.C.}, \bibinfo{author}{Purnawali, H.},
  \bibinfo{author}{Tang, C.}, \bibinfo{year}{2012}.
\newblock \bibinfo{title}{Stimulus-responsive shape memory materials: a
  review}.
\newblock \bibinfo{journal}{Materials \& Design} \bibinfo{volume}{33},
  \bibinfo{pages}{577--640}.
\bibitem[{Tan et~al.(2021)Tan, Susanto, {Anwar Ali} and Tee}]{tan2021}
\bibinfo{author}{Tan, Y.J.}, \bibinfo{author}{Susanto, G.J.},
  \bibinfo{author}{{Anwar Ali}, H.P.}, \bibinfo{author}{Tee, B.C.},
  \bibinfo{year}{2021}.
\newblock \bibinfo{title}{Progress and roadmap for intelligent self-healing
  materials in autonomous robotics}.
\newblock \bibinfo{journal}{Advanced Materials} \bibinfo{volume}{33},
  \bibinfo{pages}{2002800}.
\bibitem[{Toner and Tu(1998)}]{toner1998}
\bibinfo{author}{Toner, J.}, \bibinfo{author}{Tu, Y.}, \bibinfo{year}{1998}.
\newblock \bibinfo{title}{Flocks, herds, and schools: {A} quantitative theory
  of flocking}.
\newblock \bibinfo{journal}{Physical Review E} \bibinfo{volume}{58},
  \bibinfo{pages}{4828}.
\bibitem[{Toupin(1962)}]{toupin1962}
\bibinfo{author}{Toupin, R.}, \bibinfo{year}{1962}.
\newblock \bibinfo{title}{Elastic materials with couple-stresses}.
\newblock \bibinfo{journal}{Archive for Rational Mechanics and Analysis}
  \bibinfo{volume}{11}, \bibinfo{pages}{385--414}.
\bibitem[{Trainiti and Ruzzene(2016)}]{trainiti2016}
\bibinfo{author}{Trainiti, G.}, \bibinfo{author}{Ruzzene, M.},
  \bibinfo{year}{2016}.
\newblock \bibinfo{title}{Non-reciprocal elastic wave propagation in
  spatiotemporal periodic structures}.
\newblock \bibinfo{journal}{New Journal of Physics} \bibinfo{volume}{18},
  \bibinfo{pages}{083047}.
\bibitem[{Veenstra et~al.(2025)Veenstra, Scheibner, Brandenbourger, Binysh,
  Souslov, Vitelli and Coulais}]{veenstra2025}
\bibinfo{author}{Veenstra, J.}, \bibinfo{author}{Scheibner, C.},
  \bibinfo{author}{Brandenbourger, M.}, \bibinfo{author}{Binysh, J.},
  \bibinfo{author}{Souslov, A.}, \bibinfo{author}{Vitelli, V.},
  \bibinfo{author}{Coulais, C.}, \bibinfo{year}{2025}.
\newblock \bibinfo{title}{Adaptive locomotion of active solids}.
\newblock \bibinfo{journal}{Nature} \bibinfo{volume}{639},
  \bibinfo{pages}{935--941}.
\bibitem[{Vicsek et~al.(1995)Vicsek, Czir{\'o}k, Ben-Jacob, Cohen and
  Shochet}]{vicsek1995}
\bibinfo{author}{Vicsek, T.}, \bibinfo{author}{Czir{\'o}k, A.},
  \bibinfo{author}{Ben-Jacob, E.}, \bibinfo{author}{Cohen, I.},
  \bibinfo{author}{Shochet, O.}, \bibinfo{year}{1995}.
\newblock \bibinfo{title}{Novel type of phase transition in a system of
  self-driven particles}.
\newblock \bibinfo{journal}{Physical Review Letters} \bibinfo{volume}{75},
  \bibinfo{pages}{1226}.
\bibitem[{Vicsek and Zafeiris(2012)}]{vicsek2012}
\bibinfo{author}{Vicsek, T.}, \bibinfo{author}{Zafeiris, A.},
  \bibinfo{year}{2012}.
\newblock \bibinfo{title}{Collective motion}.
\newblock \bibinfo{journal}{Physics Reports} \bibinfo{volume}{517},
  \bibinfo{pages}{71--140}.
\bibitem[{Wang et~al.(2024)Wang, Wang, Chen, Liu, Wang, Yuan et~al.}]{wang2024}
\bibinfo{author}{Wang, J.}, \bibinfo{author}{Wang, G.}, \bibinfo{author}{Chen,
  H.}, \bibinfo{author}{Liu, Y.}, \bibinfo{author}{Wang, P.},
  \bibinfo{author}{Yuan, D.}, et~al., \bibinfo{year}{2024}.
\newblock \bibinfo{title}{Robo-matter towards reconfigurable multifunctional
  smart materials}.
\newblock \bibinfo{journal}{Nature Communications} \bibinfo{volume}{15},
  \bibinfo{pages}{8853}.
\bibitem[{Xu et~al.(2026)Xu, Wang, Wang, Liu, Liu, Zhao et~al.}]{xu2026}
\bibinfo{author}{Xu, C.}, \bibinfo{author}{Wang, S.}, \bibinfo{author}{Wang,
  H.}, \bibinfo{author}{Liu, X.}, \bibinfo{author}{Liu, Z.},
  \bibinfo{author}{Zhao, Y.}, et~al., \bibinfo{year}{2026}.
\newblock \bibinfo{title}{Mechanical non-reciprocity programmed by shear
  jamming in soft composite solids}.
\newblock \bibinfo{journal}{Nature Materials} \bibinfo{volume}{25},
  \bibinfo{pages}{117--123}.
\bibitem[{Yang et~al.(2025)Yang, Wang, Wyman, Zhai, Ma and Pasini}]{yang2025}
\bibinfo{author}{Yang, H.}, \bibinfo{author}{Wang, W.J.},
  \bibinfo{author}{Wyman, O.}, \bibinfo{author}{Zhai, W.}, \bibinfo{author}{Ma,
  L.}, \bibinfo{author}{Pasini, D.}, \bibinfo{year}{2025}.
\newblock \bibinfo{title}{Temperature-responsive multistable kirigami with
  reprogrammable multi-shape memory}.
\newblock \bibinfo{journal}{Materials Today} \bibinfo{volume}{86},
  \bibinfo{pages}{10--27}.
\bibitem[{Young et~al.(2013)Young, Scardovi, Cavagna, Giardina and
  Leonard}]{young2013}
\bibinfo{author}{Young, G.F.}, \bibinfo{author}{Scardovi, L.},
  \bibinfo{author}{Cavagna, A.}, \bibinfo{author}{Giardina, I.},
  \bibinfo{author}{Leonard, N.E.}, \bibinfo{year}{2013}.
\newblock \bibinfo{title}{Starling flock networks manage uncertainty in
  consensus at low cost}.
\newblock \bibinfo{journal}{PLoS Computational Biology} \bibinfo{volume}{9},
  \bibinfo{pages}{e1002894}.
\bibitem[{Yuan et~al.(2025)Yuan, Lejeune and Park}]{yuan2025}
\bibinfo{author}{Yuan, L.}, \bibinfo{author}{Lejeune, E.},
  \bibinfo{author}{Park, H.S.}, \bibinfo{year}{2025}.
\newblock \bibinfo{title}{Machine learning-guided design of non-reciprocal and
  asymmetric elastic chiral metamaterials}.
\newblock \bibinfo{journal}{International Journal of Solids and Structures}
  \bibinfo{volume}{318}, \bibinfo{pages}{113412}.
\bibitem[{Zechner et~al.(2020)Zechner, Nerli and Norden}]{zechner2020}
\bibinfo{author}{Zechner, C.}, \bibinfo{author}{Nerli, E.},
  \bibinfo{author}{Norden, C.}, \bibinfo{year}{2020}.
\newblock \bibinfo{title}{Stochasticity and determinism in cell fate
  decisions}.
\newblock \bibinfo{journal}{Development} \bibinfo{volume}{147},
  \bibinfo{pages}{dev181495}.
\bibitem[{Zhang et~al.(2022)Zhang, Chermprayong, Xiao, Tzoumanikas, Dams, Kay
  et~al.}]{zhang2022a}
\bibinfo{author}{Zhang, K.}, \bibinfo{author}{Chermprayong, P.},
  \bibinfo{author}{Xiao, F.}, \bibinfo{author}{Tzoumanikas, D.},
  \bibinfo{author}{Dams, B.}, \bibinfo{author}{Kay, S.}, et~al.,
  \bibinfo{year}{2022}.
\newblock \bibinfo{title}{Aerial additive manufacturing with multiple
  autonomous robots}.
\newblock \bibinfo{journal}{Nature} \bibinfo{volume}{609},
  \bibinfo{pages}{709--717}.
\bibitem[{Zhang et~al.(2021)Zhang, Mozaffari and {de Pablo}}]{zhang2021a}
\bibinfo{author}{Zhang, R.}, \bibinfo{author}{Mozaffari, A.},
  \bibinfo{author}{{de Pablo}, J.J.}, \bibinfo{year}{2021}.
\newblock \bibinfo{title}{Autonomous materials systems from active liquid
  crystals}.
\newblock \bibinfo{journal}{Nature Reviews Materials} \bibinfo{volume}{6},
  \bibinfo{pages}{437--453}.
\bibitem[{Zhang et~al.(2024)Zhang, Raymond, Lahann and
  Pena-Francesch}]{zhang2024}
\bibinfo{author}{Zhang, Z.}, \bibinfo{author}{Raymond, J.E.},
  \bibinfo{author}{Lahann, J.}, \bibinfo{author}{Pena-Francesch, A.},
  \bibinfo{year}{2024}.
\newblock \bibinfo{title}{Janus swarm metamaterials for information display,
  memory, and encryption}.
\newblock \bibinfo{journal}{Advanced Materials} \bibinfo{volume}{36},
  \bibinfo{pages}{2406149}.
\bibitem[{Zhou and Wang(2026)}]{zhou2026a}
\bibinfo{author}{Zhou, W.}, \bibinfo{author}{Wang, Y.Z.}, \bibinfo{year}{2026}.
\newblock \bibinfo{title}{Reprogrammable metamaterial robot with embodied
  versatile computation and mechanical intelligence}.
\newblock \bibinfo{journal}{Nature Communications} .

\end{thebibliography}

\end{document}